\documentclass[prc,superscriptaddress,showpacs,twocolumn,amssymb,amsmath,amsfonts,aps]{revtex4}
\usepackage[pdftex]{graphicx}
\usepackage{dcolumn}  
\usepackage{longtable}
\usepackage{bm}       
\usepackage{subfigure}
\usepackage{epstopdf}
\newcommand\captionof[1]{\def\@captype{#1}\caption}
\begin{document}
%
%
%
%
\newcommand*{\CMU}{Carnegie Mellon University, Pittsburgh, Pennsylvania 15213}
\newcommand*{\CMUindex}{1}
\affiliation{\CMU}
\newcommand*{\ANL}{Argonne National Laboratory, Argonne, Illinois 60441}
\newcommand*{\ANLindex}{2}
\affiliation{\ANL}
\newcommand*{\ASU}{Arizona State University, Tempe, Arizona 85287-1504}
\newcommand*{\ASUindex}{3}
\affiliation{\ASU}
\newcommand*{\CSU}{California State University, Dominguez Hills, Carson, CA 90747}
\newcommand*{\CSUindex}{4}
\affiliation{\CSU}
\newcommand*{\Canisius}{Canisius College, Buffalo, NY 14208}
\newcommand*{\Canisiusindex}{5}
\affiliation{\Canisius}
\newcommand*{\CUA}{Catholic University of America, Washington, D.C. 20064}
\newcommand*{\CUAindex}{6}
\affiliation{\CUA}
\newcommand*{\SACLAY}{CEA, Centre de Saclay, Irfu/Service de Physique Nucl\'eaire, 91191 Gif-sur-Yvette, France}
\newcommand*{\SACLAYindex}{7}
\affiliation{\SACLAY}
\newcommand*{\CNU}{Christopher Newport University, Newport News, Virginia 23606}
\newcommand*{\CNUindex}{8}
\affiliation{\CNU}
\newcommand*{\UCONN}{University of Connecticut, Storrs, Connecticut 06269}
\newcommand*{\UCONNindex}{9}
\affiliation{\UCONN}
\newcommand*{\ECOSSEE}{Edinburgh University, Edinburgh EH9 3JZ, United Kingdom}
\newcommand*{\ECOSSEEindex}{10}
\affiliation{\ECOSSEE}
\newcommand*{\FU}{Fairfield University, Fairfield CT 06824}
\newcommand*{\FUindex}{11}
\affiliation{\FU}
\newcommand*{\FIU}{Florida International University, Miami, Florida 33199}
\newcommand*{\FIUindex}{12}
\affiliation{\FIU}
\newcommand*{\FSU}{Florida State University, Tallahassee, Florida 32306}
\newcommand*{\FSUindex}{13}
\affiliation{\FSU}
\newcommand*{\GWU}{The George Washington University, Washington, DC 20052}
\newcommand*{\GWUindex}{14}
\affiliation{\GWU}
\newcommand*{\ECOSSEG}{University of Glasgow, Glasgow G12 8QQ, United Kingdom}
\newcommand*{\ECOSSEGindex}{15}
\affiliation{\ECOSSEG}
\newcommand*{\ISU}{Idaho State University, Pocatello, Idaho 83209}
\newcommand*{\ISUindex}{16}
\affiliation{\ISU}
\newcommand*{\INFNFR}{INFN, Laboratori Nazionali di Frascati, 00044 Frascati, Italy}
\newcommand*{\INFNFRindex}{17}
\affiliation{\INFNFR}
\newcommand*{\INFNGE}{INFN, Sezione di Genova, 16146 Genova, Italy}
\newcommand*{\INFNGEindex}{18}
\affiliation{\INFNGE}
\newcommand*{\INFNRO}{INFN, Sezione di Roma Tor Vergata, 00133 Rome, Italy}
\newcommand*{\INFNROindex}{19}
\affiliation{\INFNRO}
\newcommand*{\ORSAY}{Institut de Physique Nucl\'eaire ORSAY, Orsay, France}
\newcommand*{\ORSAYindex}{20}
\affiliation{\ORSAY}
\newcommand*{\ITEP}{Institute of Theoretical and Experimental Physics, Moscow, 117259, Russia}
\newcommand*{\ITEPindex}{21}
\affiliation{\ITEP}
\newcommand*{\JMU}{James Madison University, Harrisonburg, Virginia 22807}
\newcommand*{\JMUindex}{22}
\affiliation{\JMU}
\newcommand*{\KYUNGPOOK}{Kyungpook National University, Daegu 702-701, Republic of Korea}
\newcommand*{\KYUNGPOOKindex}{23}
\affiliation{\KYUNGPOOK}
\newcommand*{\UNH}{University of New Hampshire, Durham, New Hampshire 03824-3568}
\newcommand*{\UNHindex}{24}
\affiliation{\UNH}
\newcommand*{\NSU}{Norfolk State University, Norfolk, Virginia 23504}
\newcommand*{\NSUindex}{25}
\affiliation{\NSU}
\newcommand*{\OHIOU}{Ohio University, Athens, Ohio  45701}
\newcommand*{\OHIOUindex}{26}
\affiliation{\OHIOU}
\newcommand*{\ODU}{Old Dominion University, Norfolk, Virginia 23529}
\newcommand*{\ODUindex}{27}
\affiliation{\ODU}
\newcommand*{\RPI}{Rensselaer Polytechnic Institute, Troy, New York 12180-3590}
\newcommand*{\RPIindex}{28}
\affiliation{\RPI}
\newcommand*{\URICH}{University of Richmond, Richmond, Virginia 23173}
\newcommand*{\URICHindex}{29}
\affiliation{\URICH}
\newcommand*{\ROMAII}{Universita' di Roma Tor Vergata, 00133 Rome Italy}
\newcommand*{\ROMAIIindex}{30}
\affiliation{\ROMAII}
\newcommand*{\MOSCOW}{Skobeltsyn Nuclear Physics Institute, Skobeltsyn Nuclear Physics Institute, 119899 Moscow, Russia}
\newcommand*{\MOSCOWindex}{31}
\affiliation{\MOSCOW}
\newcommand*{\SCAROLINA}{University of South Carolina, Columbia, South Carolina 29208}
\newcommand*{\SCAROLINAindex}{32}
\affiliation{\SCAROLINA}
\newcommand*{\JLAB}{Thomas Jefferson National Accelerator Facility, Newport News, Virginia 23606}
\newcommand*{\JLABindex}{33}
\affiliation{\JLAB}
\newcommand*{\UNIONC}{Union College, Schenectady, NY 12308}
\newcommand*{\UNIONCindex}{34}
\affiliation{\UNIONC}
\newcommand*{\UTFSM}{Universidad T\'{e}cnica Federico Santa Mar\'{i}a, Casilla 110-V Valpara\'{i}so, Chile}
\newcommand*{\UTFSMindex}{35}
\affiliation{\UTFSM}
\newcommand*{\VIRGINIA}{University of Virginia, Charlottesville, Virginia 22901}
\newcommand*{\VIRGINIAindex}{36}
\affiliation{\VIRGINIA}
\newcommand*{\WM}{College of William and Mary, Williamsburg, Virginia 23187-8795}
\newcommand*{\WMindex}{37}
\affiliation{\WM}
\newcommand*{\YEREVAN}{Yerevan Physics Institute, 375036 Yerevan, Armenia}
\newcommand*{\YEREVANindex}{38}
\affiliation{\YEREVAN}
 
\newcommand*{\NOWIMPERIAL}{Imperial College London, London SW7 2AZ, UK.} 
\newcommand*{\NOWSTANFORD}{Stanford University, Stanford, CA 94305, USA.}
\newcommand*{\NOWLPSC}{LPSC-Grenoble, France.}
\newcommand*{\NOWJLAB}{Thomas Jefferson National Accelerator Facility, Newport News, Virginia 23606, USA.}
\newcommand*{\NOWLANL}{Los Alamos National Laborotory, New Mexico, NM, USA.}
\newcommand*{\NOWMINN}{University of Minnesota, Minneapolis, MN 55455, USA.}
\newcommand*{\NOWECOSSEE}{Edinburgh University, Edinburgh EH9 3JZ, UK.}
\newcommand*{\NOWWM}{College of William and Mary, Williamsburg, Virginia 23187-8795, USA.}
\newcommand*{\NOWGW}{The George Washington University, Washington, DC 20052, USA.}


\author{M.~Williams}
\altaffiliation[Current address: ]{\NOWIMPERIAL}
\affiliation{\CMU}
\author {D.~Applegate}
\altaffiliation[Current address: ]{\NOWSTANFORD}
\affiliation{\CMU}
\author {M.~Bellis} 
\altaffiliation[Current address: ]{\NOWSTANFORD}
\affiliation{\CMU}
\author {C.A.~Meyer}
\affiliation{\CMU}
\author {K. P. ~Adhikari} 
\affiliation{\ODU}
\author {M.~Anghinolfi} 
\affiliation{\INFNGE}
\author {H.~Baghdasaryan} 
\affiliation{\VIRGINIA}
\affiliation{\ODU}
\author {J.~Ball} 
\affiliation{\SACLAY}
\author {M.~Battaglieri} 
\affiliation{\INFNGE}
\author {I.~Bedlinskiy} 
\affiliation{\ITEP}
% ...

\collaboration{The CLAS Collaboration}
\noaffiliation

\date{\today}

\title{Partial wave analysis of the reaction $\gamma p \rightarrow p \omega$ 
  and the search for nucleon resonances}
%
%
\begin{abstract} 
An event-based partial wave analysis (PWA) of the reaction 
$\gamma p \rightarrow p \omega$ has been performed on a high-statistics
dataset obtained using the CLAS at Jefferson Lab for center-of-mass energies
from threshold up to 2.4~GeV. This analysis benefits from access to the 
world's first high-precision spin density matrix element measurements, 
available to the event-based PWA through the decay distribution of  
$\omega\rightarrow \pi^+\pi^-\pi^0$.  The data confirm the dominance of the 
$t$-channel $\pi^0$ exchange amplitude in the forward direction.  The dominant 
resonance contributions are consistent with the previously identified states 
$F_{15}(1680)$ and $D_{13}(1700)$ near threshold, as well as the $G_{17}(2190)$ 
at higher energies. Suggestive evidence for the presence of a $J^P=5/2^+$ state 
around $2$~GeV, a ``missing'' state, has also been found. Evidence for other states 
is inconclusive.
\end{abstract}
\maketitle
\section{\label{section:intro}INTRODUCTION}

Studying near-threshold $\omega$ photoproduction presents an
interesting opportunity to search for new baryon resonances. 
Measurements made by previous experiments have produced relatively 
high-precision cross sections at most production angles;
however, 
precise spin density matrix elements have only been measured at very forward 
angles and at higher energies
\cite{ballam-1973,clift-1977,barber-1984,battaglieri-2003,saphir-2003}.
In the near-threshold region, the only
previously published spin density matrix results, which come from the
SAPHIR collaboration, constitute a total of 8 data points in the energy range
from $\omega$ photoproduction threshold up to a center-of-mass energy,
$W$, of about $2.4$~GeV~\cite{saphir-2003}.

A number of theoretical studies have been undertaken with the goal of
extracting resonance contributions to $\omega$ photoproduction from these 
data.
All of the authors agree on the importance of contributions from 
$\pi^0$ exchange in the $t$-channel; however, discrepancies exist on the 
importance of various resonance contributions.
In the calculations of Oh {\em et al}.~\cite{oh-2001}, 
the dominant resonance contributions
are found to be from a ``missing'' $P_{13}(1910)$ state
({\em i.e.} a state predicted by the constituent quark model but not observed
experimentally) and 
from a $D_{13}(1960)$ state.
In contrast to this, Titov and Lee~\cite{titov} find the most significant
resonance contributions to be from the $D_{13}(1520)$ and $F_{15}(1680)$ 
states.
The quark model calculations made by Zhao~\cite{zhao-2001} find that the two
most important resonance contributions to $\omega$ photoproduction come from
the $P_{13}(1720)$ and $F_{15}(1680)$ states.
The $P_{11}(1710)$ and $P_{13}(1900)$ states were found to be the dominant
resonance contributions in the coupled-channel analysis of Penner and 
Mosel~\cite{penner-2002}.

All of the models mentioned above were fit solely to differential cross 
sections. A more recent analysis~\cite{penner-2005} 
also included the spin density matrix elements published by the SAPHIR 
collaboration~\cite{saphir-2003}. This work found the largest resonant 
contributions to $\omega$ photoproduction to be from the sub-threshold 
$D_{15}(1675)$ and $F_{15}(1680)$ states.
The authors noted the importance of the strong additional constraints placed
on their model by the polarization information and concluded that:
{\em there is urgent need for precise measurements of the spin density matrix 
in more narrow energy bins to pin down the reaction picture.}

Recently published results from the CEBAF Large Acceptance Spectrometer (CLAS) 
have provided such 
measurements~\cite{williams-prc}. In the center-of-mass (c.m.) energy range 
from threshold
up to $2.84$~GeV, differential cross section results were reported at
$1960$ points in $W$ and $\cos{\theta^{\omega}_{c.m.}}$. The experiment did not
use a polarized beam or a polarized target; thus, only the 
$\rho^0_{00}$, $\rho^0_{1-1}$ and $Re(\rho^0_{10})$ elements of the spin 
density matrix could be determined (the definitions of which can be found 
in~ \cite{schilling-1970}). 
These results were reported
at $2015$ points in $W$ and $\cos{\theta^{\omega}_{c.m.}}$. The increase in 
precision for $\rho^0_{MM'}$, in the energy range overlapping the SAPHIR 
results, is approximately a factor of $148$.

In this paper, we present an {\em event-based mass-independent} 
partial wave analysis (PWA) of these data, {\em i.e.} the data
are only divided into narrow c.m. energy bins. 
In each of these narrow bins, 
the spin-independent part of any resonance propagator 
-- a complex function of $W$ -- 
is approximated as a constant complex number.
This allows us to reduce model dependence in our treatment of resonances
(see Section~\ref{section:amps:resonant} for further discussion on this
topic).

The data used in our analysis were obtained using the CLAS housed in
Hall~B at the Thomas Jefferson National Accelerator Facility. 
Real photons were produced via bremsstrahlung from a $4.02$~GeV electron beam.
The momenta of the recoiling electrons were then analyzed in order to obtain 
the energy of the photons with an uncertainty of $0.1$\%~\cite{sober}.
The physics target was filled with liquid hydrogen.
The ${\omega\rightarrow \pi^+\pi^-\pi^0}$ decay was used to select the reaction
of interest.
The momenta of the charged particles ($p,\pi^+,\pi^-$) 
were determined using the CLAS detector with an uncertainty of approximately 
$0.5$\%.
The neutral $\pi^0$ was reconstructed using kinematic fitting.
More details concerning the analysis techniques can be found 
in~\cite{williams-prc,williams-thesis}.
A detailed description of the CLAS can be found in~\cite{clas-detector}.

In total, the dataset consists of 
over $10$ million signal events divided into $112$ $10$-MeV wide 
c.m. energy bins. Our primary interest is in extracting 
possible nucleon resonance contributions; thus, we have restricted our PWA 
to include only bins with $W < 2.4$~GeV. In total, $67$ c.m. energy bins were used
in the PWA (the $W = 1.955$~GeV bin was excluded due to issues with the 
normalization calculation~\cite{williams-prc}).
This work represents the first event-based PWA results on baryons 
from photoproduction data.

\section{\label{section:forms}PWA FORMULAS}

As stated above, to limit theoretical model dependence we divided our data into
10-MeV wide $W$ bins. Thus, all formulas written below are 
intended to describe data from a narrow c.m. energy range.
In all of the work that follows, $p_i,p_f,q$ and $k$ will be used for the 
initial proton, final proton, $\omega$ and photon 4-momenta. 
The $z$-axis in the overall c.m. frame, defined by $\hat{k}$, is used as the 
angular momentum quantization axis.
The Mandelstam variables are defined as
\begin{subequations}
\label{eq:mandlestam}
\begin{eqnarray}
  s &=& (p_i + k)^2 = (p_f + q)^2 \\
  t &=& (q - k)^2 = (p_i - p_f)^2 \\
  u &=& (p_i -q)^2 = (p_f - k)^2. 
\end{eqnarray}
\end{subequations}
The mass of the proton and $\omega$ are denoted as $w_p$ and $w_{\omega}$, 
respectively.

The Lorentz invariant transition amplitude, $\mathcal{M}$, of the process 
$\gamma p \rightarrow p\omega \rightarrow p\pi^+\pi^-\pi^0$, can be written as
\begin{equation}
  \label{eq:scattering-amp}
  |\mathcal{M}(\vec{\alpha},\vec{x})|^2 = \sum\limits_{m_i,m_{\gamma},m_f}
  \left|\sum\limits_a \mathcal{A}_{m_i,m_{\gamma},m_f}^a(\vec{\alpha},\vec{x})\right|^2,
\end{equation}
where $m_i,m_{\gamma},m_f$ are the initial proton, incident photon and final 
proton spin projections,
$\mathcal{A}^a_{m_i,m_{\gamma},m_f}$ are the partial wave amplitudes
(the form of which is discussed in Section~\ref{section:amps}),
$\vec{x}$ denotes the complete set of kinematic variables describing the 
reaction
and $\vec{\alpha}$ are the unknown parameters to be determined by the fit.
We denote the detector acceptance by $\eta(\vec{x})$ and the phase-space
volume as $d\Phi(\vec{x}) = \phi(\vec{x})d\vec{x}$.
A more detailed description of the work presented in this section is given
in~\cite{williams-thesis}.

\subsection{\label{section:forms:likelihood}Likelihood}

All of our fits are event-based; thus, the data were only binned in
$W$. To obtain estimators for the unknown parameters, $\hat{\alpha}$,
we employ the extended unbinned maximum likelihood method.
The work detailed in this section is based on that of Chung~\cite{chung}; 
however, the normalizations we have developed differ from his work.
The likelihood function is defined as
\begin{equation}
  \label{eq:likelihood}
  \mathcal{L} = \left(\frac{\bar{n}(\vec{\alpha})^n}{n!}
  e^{-\bar{n}(\vec{\alpha})}\right) 
  \prod\limits_i^n \mathcal{P}(\vec{\alpha},\vec{x}_i),
\end{equation}
where the term in parentheses is the Poisson probability of obtaining $n$ 
events given the expected number  $\bar{n}(\vec{\alpha})$ 
(the calculation of which is 
discussed below), $\vec{x}_i$ represents the complete set of kinematic 
variables of the $i^{th}$ event and $\mathcal{P}(\vec{\alpha},\vec{x})$ is 
the probability density function given by
\begin{equation}
  \label{eq:pdf}
  \mathcal{P}(\vec{\alpha},\vec{x}_i) 
  = \frac{|\mathcal{M}(\vec{\alpha},\vec{x}_i)|^2 
    \eta(\vec{x}_i) \phi(\vec{x}_i)}
  {\int |\mathcal{M}(\vec{\alpha},\vec{x})|^2 
    \eta(\vec{x}) \phi(\vec{x}) d\vec{x}}.
\end{equation}
>From left to right, Eq.~(\ref{eq:pdf}) accounts for the relative strength of
the transition amplitude, the detection probability and the available phase 
space for the $i^{th}$ event. Calculation of the denominator, which normalizes 
the probability density function, is discussed below.
The estimators $\hat{\alpha}$ are then found by maximizing $\mathcal{L}$.

\subsection{\label{section:forms:norm}Normalization}

The expected number of signal events for a given set of parameters is given by
\begin{equation}
  \label{eq:nbar-def}
  \bar{n}(\vec{\alpha}) = 
  \frac{\mathcal{T}(s)(2\pi)^4}{8(s-w_p^2)}\int 
  |\mathcal{M}(\vec{\alpha},\vec{x})|^2 \eta(\vec{x})d\Phi(\vec{x}),
\end{equation}
which includes the average over initial spin states. 
\begin{equation}
  \mathcal{T}(s) = \frac{\mathcal{F}(s)\rho_{targ}\ell_{targ}N_A b}{A_{targ}}
\end{equation}
is the ``target factor'' obtained from the 
target density, $\rho_{targ}$, length, $\ell_{targ}$, and atomic number,
$A_{targ}$; along with Avogadro's number, $N_A$; the branching fraction of 
$\omega\rightarrow\pi^{+}\pi^{-}\pi^{0}$, $b$; and the
integrated photon flux in each $W$ bin, $\mathcal{F}(s)$. 

The integral in Eq.~(\ref{eq:nbar-def}) must be done numerically due to the 
lack of
an analytic expression for the detector acceptance. Monte Carlo events were 
generated in each $W$ bin according to $\gamma p \rightarrow p\omega$
(the $\omega$ mass was generated according to a Breit-Wigner distribution),
$\omega \rightarrow \pi^+\pi^-\pi^0$ phase space 
and then run through a 
GEANT-based detector simulation package (discussed in detail in 
\cite{williams-thesis,williams-prc}). 
This procedure simulates 
the acceptance of the detector by rejecting events that would not have 
survived the data analysis, {\em i.e.} for each generated event, 
the acceptance factor $\eta(\vec{x}_i) = 0$ or $1$.
The integral can then be approximated by
\begin{eqnarray}
  \label{eq:nbar-integral}
  \int |\mathcal{M}(\vec{\alpha},\vec{x})|^2 \eta(\vec{x}) d\Phi(\vec{x}) 
  \hspace{0.2\textwidth}
  \nonumber \\
  \hspace{0.2\textwidth}
  \approx \frac{\int d\Phi(\vec{x})}{n_{gen}} 
  \sum\limits_i^{n_{acc}} |\mathcal{M}(\vec{\alpha},\vec{x}_i)|^2,
\end{eqnarray}
where $n_{gen}(n_{acc})$ is the number of generated (accepted) Monte 
Carlo events and
\begin{equation}
  \label{eq:phase-space}
  \int d\Phi(\vec{x}) =  
  \frac{\left[(s - (w_p + w_{\omega})^2)(s - (w_p - w_{\omega})^2)
      \right]^{1/2}}
       {4(2\pi)^5 s}
\end{equation}
is the volume of the 2-body $p\omega$ phase space (the $3\pi$ phase-space 
volume is factored into the normalization of the $\omega$ decay amplitude).

Using Eqs.~(\ref{eq:nbar-integral}) and (\ref{eq:phase-space}),
Eq.~(\ref{eq:nbar-def}) can be rewritten as
\begin{eqnarray}
\bar{n}(\vec{\alpha}) = 
  \frac{\left[(s - (w_p + w_{\omega})^2)(s - (w_p - w_{\omega})^2)
      \right]^{1/2}}
       {64 \pi s (s - w_p^2)}
       \nonumber \\
       \times \frac{\mathcal{T}(s)}{n_{gen}}
       \sum\limits_i^{n_{acc}} |\mathcal{M}(\vec{\alpha},\vec{x}_i)|^2.
\end{eqnarray}
This normalization allows us to use physical coupling constants 
in our event-based fits, {\em i.e.} it allows us 
to put our parameters on an absolute scale. Thus, our normalization scheme
permits direct theoretical input.

\subsection{\label{section:forms:log-L}Log likelihood}

Due to the monotonically increasing nature of the natural logarithm, the 
likelihood, $\mathcal{L}$, defined in Eq.~(\ref{eq:likelihood})  can be 
maximized by minimizing
\begin{equation}
  \label{eq:log-L-def}
  -\ln{\mathcal{L}} = -n \ln{\bar{n}(\vec{\alpha})} + \ln{n!} 
  + \bar{n}(\vec{\alpha}) 
  - \sum\limits_i^n \ln{\mathcal{P}(\vec{\alpha},\vec{x}_i)},
\end{equation}
which, using Eqs.~(\ref{eq:pdf}) and (\ref{eq:nbar-def}), can be rewritten 
as
\begin{eqnarray}
  \label{eq:log-L-def-v2}
  -\ln{\mathcal{L}} = \ln{n!} + \bar{n}(\vec{\alpha})
   + n\ln{\frac{8(s-w_p^2)}{\mathcal{T}(s)(2\pi)^4}} \hspace{0.4in}
  \nonumber \\
  \hspace{0.5in}
  -\sum\limits_i^n \ln{|\mathcal{M}(\vec{\alpha},\vec{x}_i)|^2 \eta(\vec{x}_i) \phi(\vec{x}_i)}.
\end{eqnarray}
Neglecting terms that do not depend on the parameters, we 
can then rewrite Eq.~(\ref{eq:log-L-def-v2}) as follows:
\begin{equation}
  \label{eq:log-L}
   -\ln{\mathcal{L}} 
   = -\sum\limits_i^n \ln{|\mathcal{M}(\vec{\alpha},\vec{x}_i)|^2}
   + \bar{n}(\vec{\alpha}) + const.
\end{equation}
We note here that for any set of estimators that minimize 
Eq.~(\ref{eq:log-L}), the expected number of events is 
$\bar{n}(\hat{\alpha})=n$.

\subsection{\label{section:forms:bkgd}Handling background}

To accurately extract partial wave contributions to $\omega$ photoproduction,
background events, {\em i.e.} non-$\omega$ events, must be separated from
the signal in a way that preserves all kinematic correlations. The 
method we applied to our data, described in detail in 
\cite{williams-prc,bkgd-preprint,williams-thesis},
assigned each event a signal quality factor, or $Q$-factor. This background was
assumed to be non-interfering. Following our previous work~\cite{bkgd-preprint}, we 
can rewrite Eq.~(\ref{eq:log-L}) using these $Q$-factors as
\begin{equation}
  \label{eq:log-L-Q}
   -\ln{\mathcal{L}} 
   = -\sum\limits_i^n Q_i \ln{|\mathcal{M}(\vec{\alpha},\vec{x}_i)|^2}
   + \bar{n}(\vec{\alpha}) + const,
\end{equation}
where $Q_i$ is the $Q$-factor for the $i^{th}$ event.
Thus, the $Q$-factors are used to weight each event's contribution to the 
likelihood. We also note that in the literature, the $t$- and $u$-channel
contributions are often referred to as \emph{background}.  This \emph{theoretical}
background is not to be confused with the experimental background discussed 
here. In this analysis, the theoretical backgrounds were allowed
to interfere with the $s$-channel amplitudes in our PWA.

\section{\label{section:amps}PWA AMPLITUDES}

\begin{figure}
  \centering
  \subfigure[]{
    \includegraphics[width=0.22\textwidth]{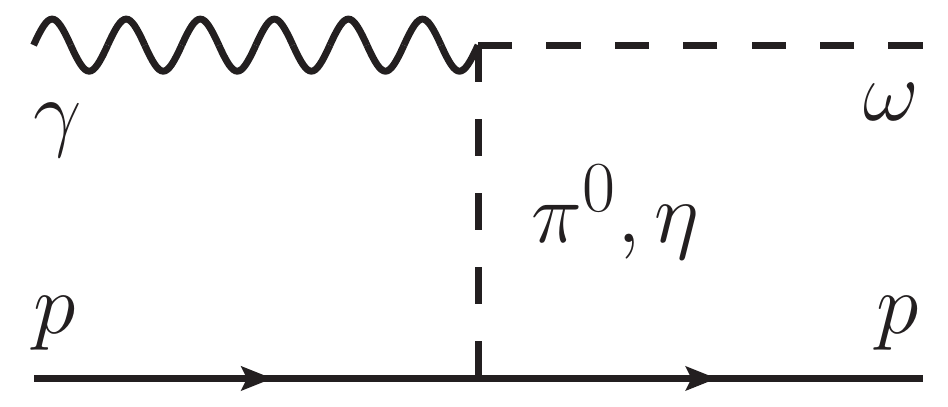}
  }
  \subfigure[]{
    \includegraphics[width=0.22\textwidth]{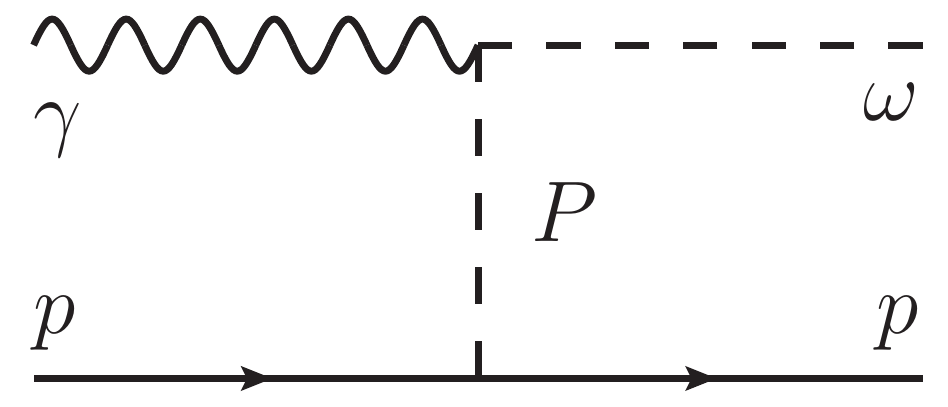}
  }
  \\
  \subfigure[]{
    \includegraphics[width=0.22\textwidth]{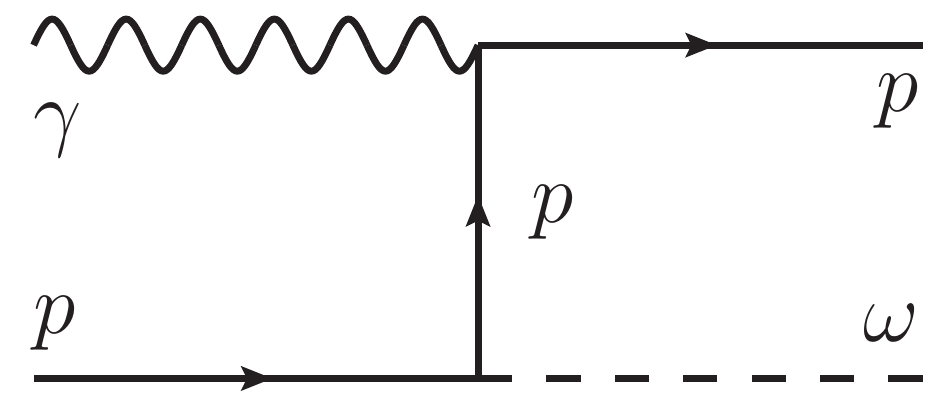}
  }
  \subfigure[]{
    \includegraphics[width=0.22\textwidth]{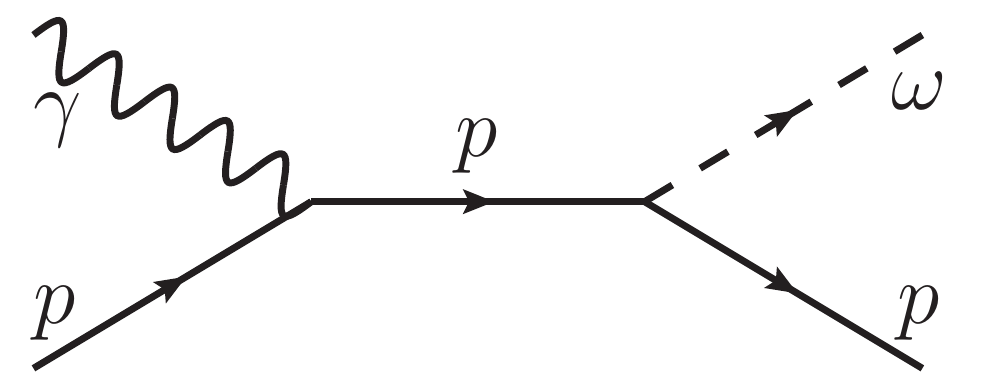}
  }
  \\
  \subfigure[]{
    \includegraphics[width=0.22\textwidth]{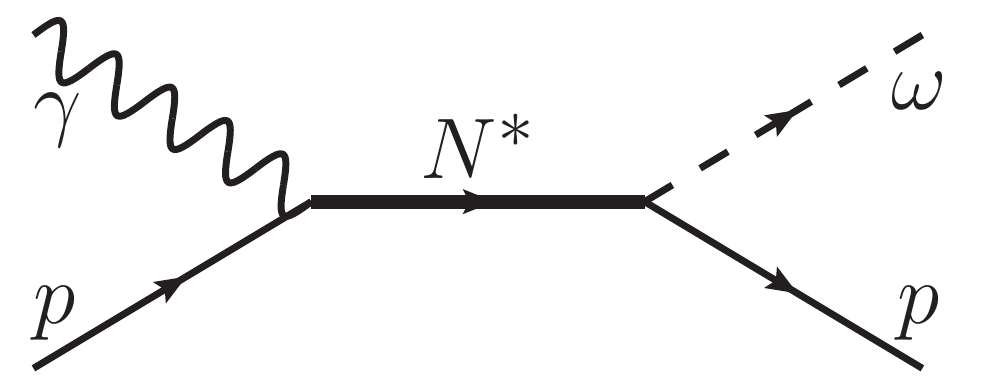}
  }
  \caption[]{\label{fig:feynman-diagrams}
    Feynman diagrams for the amplitudes used in our analysis. The  images
    were produced using the JaxoDraw package~\cite{jaxodraw}.
  }
\end{figure}

The choice of which amplitudes to include to describe the data is partially
motivated by experimental measurements. 
The $\omega$ photoproduction cross section is known to have a strong forward
peak, even at near-threshold energies~\cite{williams-prc}.
At higher energies, the cross section develops a rather pronounced backwards
peak as well~\cite{williams-prc}.
These features are typically associated with meson and nucleon exchanges in 
the $t$- and $u$-channel, respectively.
The recent CLAS data also possess a number of features in the cross sections
and spin density matrix elements suggestive of resonance 
contributions~\cite{williams-prc}.
Thus, it would seem that 
$s$-, $t$- and $u$-channel amplitudes may be required to fully describe the 
data (see Fig.~\ref{fig:feynman-diagrams}).
The formalism used to construct these amplitudes is described in detail
elsewhere~\cite{williams-thesis}, below we simply give an overview of the 
different types of amplitudes used in our analysis. All of these amplitudes 
were computed using the {\tt qft++} package~\cite{qft-software}, which 
performs numerical computations of quantum field theory expressions.

\subsection{\label{section:amps:omega-decay}$\omega\rightarrow\pi^+\pi^-\pi^0$}

The $\omega \rightarrow \pi^+ \pi^- \pi^0$ amplitude can be written in terms of
the isovectors, $\vec{I}_{\pi}$, and the 4-momenta, $p_{\pi}$, of the pions,
along with the $\omega$ 4-momentum ($q$), polarization ($\epsilon$), and spin projection, 
$m_{\omega}$ as
\begin{eqnarray}
  \label{eq:omega-decay-amp}
  \mathcal{A}^{m_{\omega}}_{\omega \rightarrow \pi^+ \pi^- \pi^0} \propto 
  \left((\vec{I}_{\pi^+} \times \vec{I}_{\pi^0}) \cdot \vec{I}_{\pi^-} \right)
  \hspace{0.1\textwidth}\nonumber \\
  \hspace{-0.1in}
  \times \epsilon_{\mu\nu\alpha\beta}p_{\pi^+}^{\nu}p_{\pi^-}^{\alpha}
  p_{\pi^0}^{\beta} \epsilon^{\mu}(q,m_{\omega}),
\end{eqnarray}
which is fully symmetric under interchange of the three pions. For this reaction, where 
all final states contain $\omega\rightarrow\pi^{+}\pi^{-}\pi^{0}$, the isovector triple 
product simply contributes a factor to the global phase of all amplitudes.
In the $\omega$ rest frame, Eq.~(\ref{eq:omega-decay-amp}) simplifies to
\begin{equation}
    \mathcal{A}^{m_{\omega}}_{\omega \rightarrow \pi^+ \pi^- \pi^0} \propto
    \left( \vec{p}_{\pi^+} \times \vec{p}_{\pi^-} \right)
    \cdot  \vec{\epsilon}(m_{\omega}),
\end{equation}
which is the standard non-relativistic result~\cite{zemach}.

\subsection{\label{section:amps:non-resonant}$t$- and $u$-channel}

Previous studies of forward $\omega$ photoproduction data have shown 
that the reaction is dominated at low energies by pion exchange and at 
higher energies by diffractive processes, {\em i.e.} Pomeron 
exchange (see, {\em e.g.},~\cite{ballam-1973}). 
We have chosen to use the non-resonant 
terms included in the model of Oh, Titov and Lee (OTL)~\cite{oh-2001} in our 
partial wave analysis. The OTL Pomeron exchange amplitude follows
the work of Donnachie and Landshoff~\cite{donnachie} with the unknown 
parameters fixed by fitting to high energy vector meson cross section
data.

The OTL model also includes pseudoscalar meson exchange amplitudes obtained
from the following Lagrangians:
\begin{subequations}
\begin{eqnarray}
  \mathcal{L}_{\phi pp} &=& -i g_{\phi pp} \bar{\psi} \gamma^5 \psi \phi \\
  \mathcal{L}_{\gamma\phi\omega} &=& e\frac{g_{\gamma\phi\omega}}{w_{\omega}} 
  \epsilon^{\mu\nu\alpha\beta}\partial_{\mu}\omega_{\nu}\partial_{\alpha}
   A_{\beta} \phi,
\end{eqnarray}
\end{subequations}
where $\phi=(\pi,\eta)$, $A_{\mu}$ and $\psi$ denote the pseudoscalar, photon 
and proton fields, respectively. The vertices in these amplitudes were 
dressed using form factors of the type
\begin{equation}
  F(t,\Lambda) = \frac{\Lambda^2 - w_{\phi}}{\Lambda^2 - t},
\end{equation}
where $\Lambda$ is the cutoff parameter for the interaction and $w_{\phi}$ is
the mass of the exchanged particle.
The larger mass and weaker coupling constants of the $\eta$ suppress its 
contribution relative to that of the pion.

Nucleon pole terms were obtained from the Lagrangians
\begin{subequations}
\begin{eqnarray}
  \mathcal{L}_{\gamma pp} &=& -e \bar{\psi} (\gamma^{\mu} - \frac{\kappa_p}
	  {2m_p}
  \sigma^{\mu\nu}\partial_{\nu})A_{\mu} \psi \\
  \mathcal{L}_{\omega pp} &=& -g_{\omega pp} \bar{\psi}(\gamma^{\mu} 
  - \frac{\kappa_{\omega}}{2m_p} \sigma^{\mu\nu}\partial_{\nu})\omega_{\mu}
  \psi.
\end{eqnarray}
\end{subequations}
The form factor
\begin{equation}
  F_N(x) = \frac{\Lambda_{N}^4}{\Lambda_N^4 - (x - w_p^2)^2},
\end{equation}
where $x = (s,u)$, was included to dress the corresponding vertices. 
The amplitudes were 
also modified to preserve gauge invariance.
The details concerning these modifications, along with the values of the 
parameters used in the model can be found in~\cite{oh-2001}.

In the near-threshold region, the high precision spin density matrix results
published by CLAS confirm the dominance of $t$-channel $\pi^0$ exchange in the
forward direction; however, at higher energies
the existing theoretical models fail to reproduce the CLAS 
data~\cite{williams-prl}.
The Pomeron amplitudes are able to describe the energy dependence of the 
forward cross section, but fail to adequately describe the spin density 
matrix elements.
The unknown parameters present in the nucleon exchange amplitudes can be 
modified to describe the backward-angle data at higher energies if some 
assumptions, the reliability of which are unknown, are 
made~\cite{williams-prl}.

Our analysis is restricted to the energy range from threshold up to 2.4~GeV. 
For c.m. energies below 2~GeV, the $\pi^0$ exchange 
amplitude dominates the $t$-channel contributions. 
In the higher energy range used in
our analysis, the $\pi^0$ contribution is still substantially larger than that
of the Pomeron. Thus, the deficiencies in the Pomeron amplitudes (discussed above) 
should not greatly affect our PWA results. For this reason, we have chosen to use the OTL 
$t$-channel terms with the parameter values obtained in that analysis~\cite{oh-2001}.
Due to the unreliability of the assumptions under which the nucleon exchange
parameters were determined, we have decided not to include these amplitudes
in our analysis, {\em i.e.} we do not include any $u$-channel terms in our
fits.
The effect of this choice on our conclusions was found to
be negligible (see Section~\ref{section:sys}).

\subsection{\label{section:amps:resonant}Resonant waves}

The formalism used to construct our resonant amplitudes is described fully
in~\cite{williams-thesis}. It involves the use of relativistic tensor
operators and is similar to the framework employed by 
Anisovich {\em et al.}~\cite{anisovich}.

As discussed above, we do not impose resonant-like shapes on our $s$-channel
waves. Instead, we divide our data into narrow c.m. energy bins.
In each of these bins, the mass-dependence of an $s$-channel wave with spin $J$
and parity $P$, which we will denote $\mathcal{R}_{J^P}(s)$, 
is approximated by a constant complex number:
\begin{equation}
\mathcal{R}_{J^P}(s) \approx \sum\limits_b r^b_{J^P} e^{i\phi^b_{J^P}}
\Theta(\delta - |\sqrt{s} - W_b|),
\end{equation}
where the sum is over the 10-MeV wide c.m. energy bins, $r^b_{J^P},\phi^b_{J^P}$ 
are the strength and phase of the mass dependence in each bin, respectively,
and $\delta \equiv 5$~MeV is the maximum distance in any bin from the centroid
$W_b$. The resonant waves then enter into our fits according to
\begin{equation}
  \label{eq:resonance}
  \mathcal{A}^{J^P,LS_i,LS_f}_{m_i,m_{\gamma},m_f} = 
  g^{J^P}_{LS_i} g^{J^P}_{LS_f} \mathcal{R}_{J^P}(s) 
  A^{J^P,LS_i,LS_f}_{m_i,m_{\gamma},m_f},
\end{equation}
where $LS_{i(f)}$ are the angular momentum quantum numbers of the 
initial (final) state, $g^{J^P}_{LS_i(f)}$ are the unknown coupling
constants to these states and $A$ is the covariant amplitude obtained using 
the formalism described in~\cite{williams-thesis}.

The values extracted for each $s$-channel wave's $r^b_{J^P}$ and $\phi^b_{J^P}$
parameters can be used to search for evidence of nucleon resonance
contributions in that wave. 
For every fit iteration run in each $W$ bin (multiple iterations are run to 
alleviate problems caused by local minima), 
these parameters are started at random values that include the entire 
physically allowed range of the parameter. 
For example, $\phi^b_{J^P}$ is started randomly in the range $[0,2\pi)$.
Estimators for the parameters are then found by maximizing the likelihoods 
independently in each $W$ bin. In this way, the mass dependence of the waves
is extracted unbiasedly. If resonant-like features are found in the cross sections 
and phase motion of the $s$-channel waves, then this is very strong evidence that 
resonances do contribute to the scattering amplitude. 

If the strength observed in an $s$-channel wave is due to a single resonant 
state, then $\mathcal{R}_{J^P}(s)$ should be (at least qualitatively) 
described by a constant-width Breit-Wigner line shape of the form:
\begin{equation}
  \label{eq:bw}
  BW(s) = \frac{w\Gamma}{s - w^2 + iw\Gamma},
\end{equation}
where $w$ and $\Gamma$ denote the mass and width of the state, respectively.
If, however, the strength in the wave is caused by multiple 
resonant states or from a non-resonant process, then the use of a Breit-Wigner
line shape is not valid.
The line shape given in Eq.~(\ref{eq:bw}) neglects the kinematics and dynamics
of the mass dependence of the resonance. This can be an issue, {\em e.g.}, 
near threshold.  This will be addressed below.

\section{\label{section:results}RESULTS}

Before examining our results, it is important to reiterate the goals of our
analysis. We are attempting to extract strong resonance contributions to 
$\omega$ photoproduction in a least model-dependent way.
We do not enforce resonance shapes on the mass dependence of our partial waves.
Instead, we first extract the strengths and phase motion of our partial waves 
independently in each $W$ bin.  This stage will be referred to as a partial wave
extraction (PWE). The second stage involves comparing the results of the PWE to
what is expected from resonances in a mass-dependent fit (MDF).  The PWE's are 
performed using unbinned extended maximum likelihood fits to the data in each 
$W$ bin. The MDF's are simply $\chi^2$ fits to the phase differences obtained 
between partial waves in a given $W$ region.

We are not looking to build a complete model of $\omega$ 
photoproduction, {\em i.e.} we do not claim that the fits discussed below 
contain all of the amplitudes that contribute to this reaction. 
Because of this, we do not expect the physical observables extracted from
our fits to provide perfect descriptions of our measurements; however, the
descriptions in many kinematic regions are very good. 
Finally, we are not attempting to extract all resonance contributions to
$\omega$ photoproduction, only the most significant.

\subsection{\label{section:results:scans}Choice of wave sets}

It is important to have a systematic method for selecting wave sets. 
As discussed in Section~\ref{section:amps:non-resonant}, all of the wave sets
used in our analysis contain the OTL $t$-channel terms 
(which contain no free parameters) and no $u$-channel terms. 
Systematic studies show that the effects on extracted resonance parameters due to the
choice of the non-resonant model are small (see Section~\ref{section:sys}).

The mass-independent nature of our procedure, {\em i.e.} the bin-to-bin freedom
of the resonance parameters, makes the use of smaller wave sets advantageous.
For this reason, we began our wave selection process by scanning the entire
energy range of interest, $1.72$~GeV${<W <}2.4$~GeV, using the OTL 
$t$-channel terms along with waves from a single spin-parity, $J^P$.
The goal of this scan was to identify (possible) energy ranges where waves of
a given $J^P$ perform significantly better than waves of any other 
spin-parity. This
information alone does not constitute evidence of resonance production;
however, it can serve as a guide as to which waves are more likely to 
contribute strongly to $\omega$ photoproduction. 

Given a pair of fits run with different wave sets,
the difference in the log likelihoods obtained from the fits,  
$\Delta\ln{\mathcal{L}} \equiv \ln{\mathcal{L}_a} - \ln{\mathcal{L}_b}$, 
can be used to quantitatively determine which fit best describes the data.
If $\Delta\ln{\mathcal{L}} > 0$, then wave set $a$ provides a better description
of the data than wave set $b$, while $\Delta\ln{\mathcal{L}} < 0$ implies the converse is true.

\begin{figure}[h!]
\begin{center}
  \includegraphics[width=0.49\textwidth]{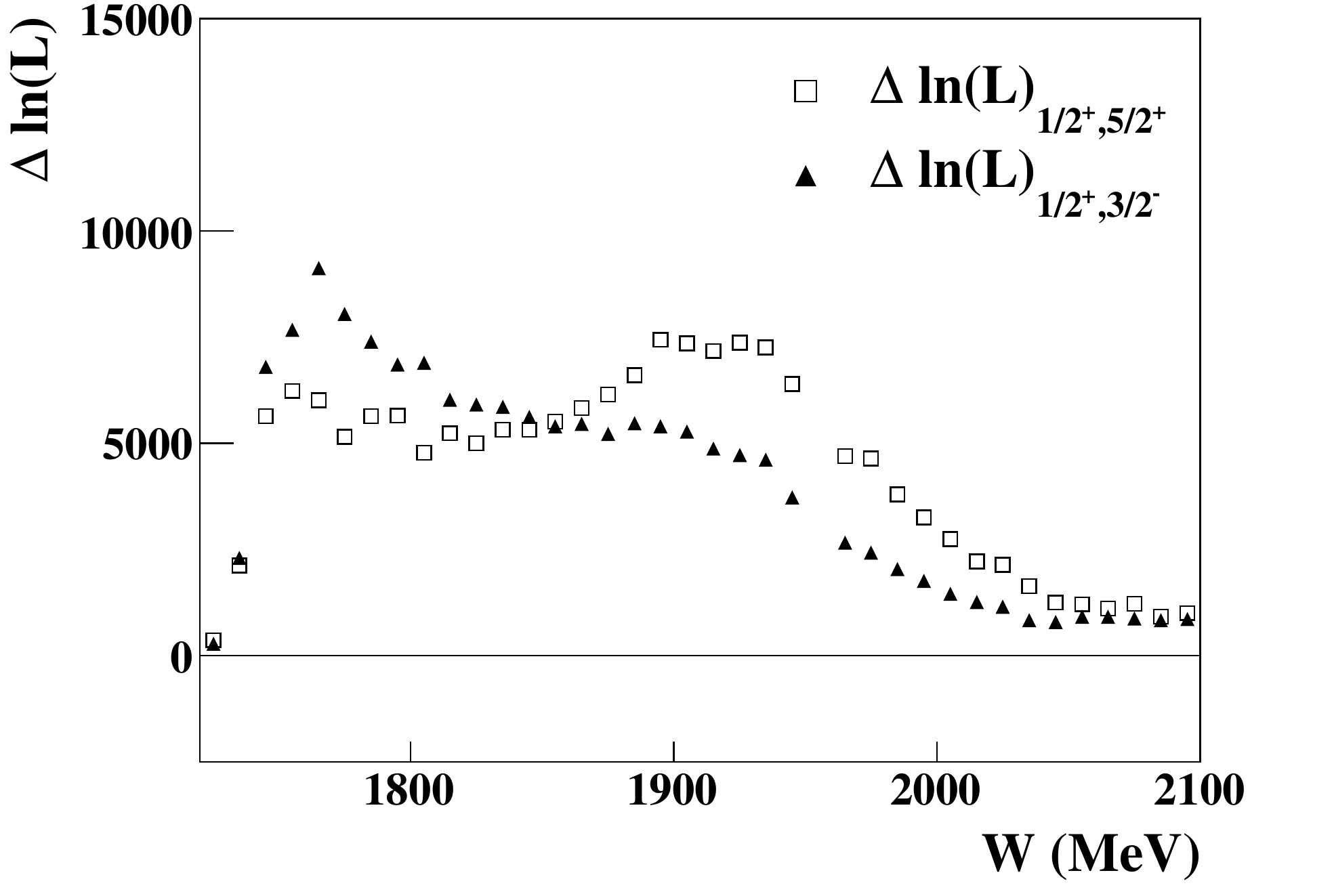}
\caption[]{
  $\Delta\ln{\mathcal{L}}$ vs $W$(MeV):
  Example likelihood differences from $s$-channel scans. Each fit contained 
  the locked OTL $t$-channel terms, along with a single $s$-channel wave. 
  Shown are 
  ${\Delta\ln{\mathcal{L}}_{1/2^+,5/2^+} = \ln{\mathcal{L}}_{5/2^+} - \ln{\mathcal{L}}_{1/2^+}}$ 
  (open squares) and 
  ${\Delta\ln{\mathcal{L}}_{1/2^+,3/2^-} = \ln{\mathcal{L}}_{3/2^-} - \ln{\mathcal{L}}_{1/2^+}}$ (closed triangles).
  See text for details and discussion.
}
\label{fig:dlogL-example}
\end{center}
\end{figure}

Figure~\ref{fig:dlogL-example} shows two examples comparing 
the likelihood differences between fits with different $s$-channel waves.
The $\Delta\ln{\mathcal{L}}$ quantities are shown for two separate fits,
one with $J^P = 5/2^+$ and one with $J^P = 3/2^-$, each of which is compared to
a fit with $J^P = 1/2^+$.
>From threshold up to $W \sim 1.85$~GeV, the fit with
$J^P = 3/2^-$ is clearly the best, while in the energy range 
1.85~GeV${<W<}$2~GeV the preferred wave is $J^P = 5/2^+$.
It is also clear in Fig.~\ref{fig:dlogL-example} that both the $J^P = 3/2^-$
and  $J^P = 5/2^+$ waves provide better descriptions of our data than the
$J^P = 1/2^+$ wave in this energy range.

Similar fits were run using any single $s$-channel wave 
with $J \leq 9/2$ of both
parities. In the region from threshold up to $W \sim 1.85$~GeV, the
$J^P = 3/2^-$ wave was found to provide a better description of our data than
any other wave. Similarly, in the energy range $1.85$~GeV${<W<}$2~GeV, 
the
$J^P = 5/2^+$ wave was found to provide the best description.

Scans were also performed using two $s$-channel waves and the OTL $t$-channel
terms. The quantity $\Delta\ln{\mathcal{L}}$ can also be used in these fits to
determine which wave sets best describe our data.
In the ${W< 2}$~GeV region, the best fit was obtained using
the $s$-channel waves with $J^P = 3/2^-,5/2^+$. 
This wave set had the best likelihood in 
every bin in this energy range (typically by a large amount). 
Given the results of the single wave $s$-channel scans discussed above, this
is not a surprising result.
Above 2~GeV, the preferred wave set consisted of the $J^P = 5/2^+,7/2^-$ waves, 
along
with the $t$-channel terms. As in the lower energy range, this wave set had
the best likelihood in every energy bin for $W>2$~GeV.

The results presented for the waves below were not affected by our
choice of wave set; however, as the number of waves was increased so did the 
noise. For this reason, we have chosen to present the results from fits
with at most three $s$-channel waves. See Section~\ref{section:sys} for more
discussion on fits with a larger number of waves.

\subsection{\label{section:results:threshold}Fit I: the near-threshold region}

Our preliminary $s$-channel scans showed that the best fit using two $s$-channel waves, 
along with the OTL $t$-channel terms,  in the energy range $1.72$~GeV~${<W<}$~$2$~GeV is 
obtained using ${J^P = 3/2^-,5/2^+}$.  To extract any possible resonance contributions in 
these waves, event-based PWE fits were run using the locked OTL $t$-channel terms along 
with $J^P = 3/2^-,5/2^+$ $s$-channel waves parametrized according to Eq.~(\ref{eq:resonance}). 
In each energy bin, multiple fit iterations were run using random starting values for
the parameters; the results presented below are always from the fit with the best likelihood. 

\subsubsection{\label{section:results:threshold:cs_and_dphase}Cross sections
  and phase motion}

\begin{figure}[h!]
\begin{center}
  \includegraphics[width=0.50\textwidth]{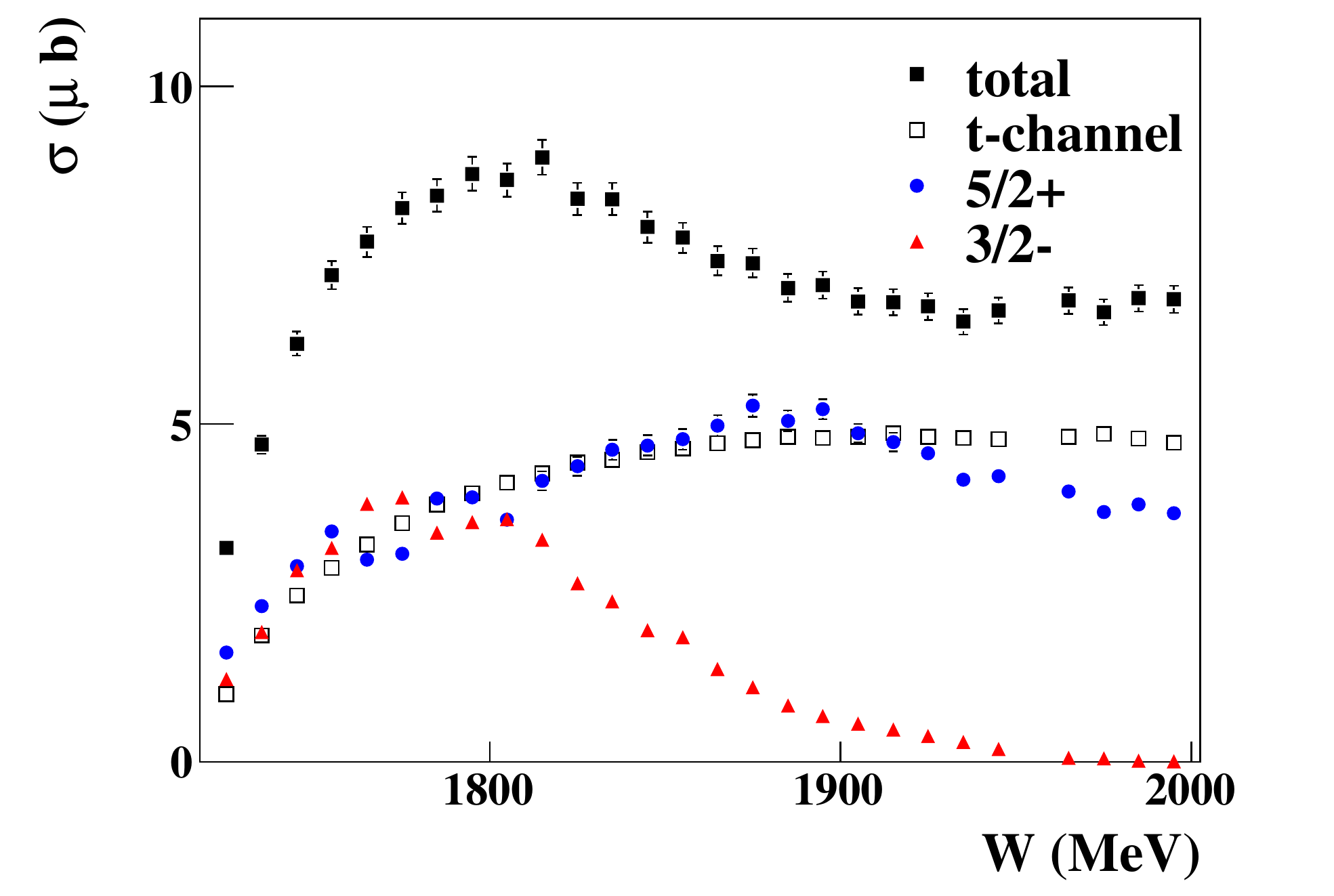}
\caption[]{
  (Color Online)
  Results from Fit I: 
  $\sigma (\mu b)$ vs $W$(MeV): Total cross sections from all of the waves 
  included
  in the fit (filled squares), only $t$-channel waves (open squares), only 
  $J^P = 5/2^+$ waves (circles) and only $J^P = 3/2^-$ waves (triangles). The
  cross sections extracted for both $s$-channel waves are consistent with 
  near/sub-threshold resonances. The error bars are purely statistical.
}
\label{fig:thresh:sigma}
\end{center}
\end{figure}

The strength and phase of each $s$-channel wave were completely free to vary in
each energy bin, {\em i.e.} they were fit independently. The cross sections extracted 
for the $s$-channel waves are consistent with either near or sub-threshold resonance 
states (see Fig.~\ref{fig:thresh:sigma}). The PDG lists two states in these waves consistent 
with this hypothesis:
(1) the 4-star $F_{15}(1680)$, which has a well known very large coupling to 
$\gamma p$;
(2) the 3-star $D_{13}(1700)$, which is currently rated as having only a 2-star
coupling to $\gamma p$. We note here that the masses of the states cannot be 
(precisely) estimated by simply examining the cross sections due to threshold 
suppression effects.

\begin{figure}[h!]\centering
 \includegraphics[width=0.50\textwidth]{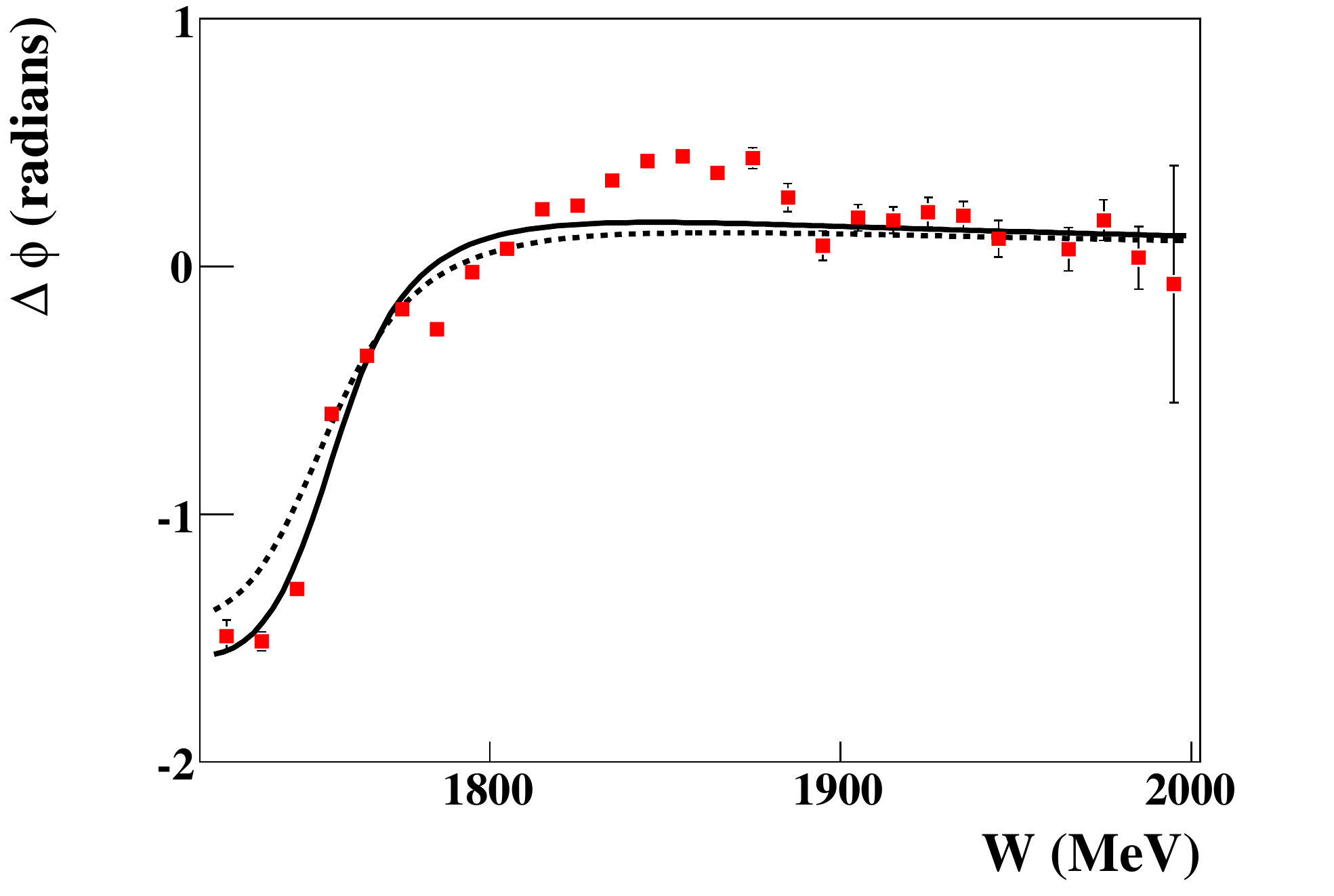}
\caption[]{
  Results from Fit I: 
  $\Delta\phi = \phi_{3/2^-} - \phi_{5/2^+}$(radians) vs $W$(MeV): 
  The dashed line
  was fit using constant width Breit-Wigner distributions requiring the parameters to 
  be within the limits quoted by the PDG for the 4-star $F_{15}(1680)$ and 
  3-star $D_{13}(1700)$. 
  The solid line was fit allowing the $3/2^-$ parameters to vary freely,
  the results are listed in the text. 
  The error bars are purely statistical.
}
\label{fig:thresh:dphase}
\end{figure}

Figure \ref{fig:thresh:dphase} shows the phase motion between the two
$s$-channel waves obtained from the PWE fits. The phase differences were then
fit in a MDF using the constant width Breit-Wigner line shapes of the form 
given in Eq.~(\ref{eq:bw}). We chose not to use mass-dependent widths in the 
Breit-Wigner line shapes  (despite the proximity to $p\omega$ threshold) since 
the $F_{15}(1680)$ (and perhaps the $D_{13}(1700)$ as well) is below threshold,
which introduces model dependence in a single-channel analysis.  For this 
reason, extracting precise values for the resonance parameters may not be 
possible; however, the use of Eq.~(\ref{eq:bw}) is sufficient to provide
evidence for the presence of known PDG states in our data.


In principle, our MDF fits could have been made to the partial-wave intensities
as well as the phase difference. In order to do this, form factors would need to be introduced 
at both the production and decay vertices. While these form factors do not strongly influence 
the phase difference, they are very important for describing the shapes of the cross sections.
The intensities of the partial waves (in this fit, and in the following sections) are qualitatively 
consistent with the expected resonance shapes. Obtaining good quantitative agreement requires 
the extra degrees of freedom introduced by the form factors; however, including these factors 
also introduces additional model dependence. Thus, we have decided to only fit the phase 
difference --- which, as noted, is nearly independent of the form factors.

Our results are, qualitatively, in good agreement with 
those expected from the PDG states mentioned above.
The dashed-line in Fig.~\ref{fig:thresh:dphase} was fit requiring all 
parameters to be within the limits quoted by the PDG for the 
$F_{15}(1680)$ and $D_{13}(1700)$. There is a minor
discrepancy in the near-threshold bins.
The parameters of the $D_{13}(1700)$ are not as well known as those of the
$F_{15}(1680)$; thus, we also performed a MDF allowing the $3/2^-$
parameters to vary freely. This fit resulted in a mass of $1754$~MeV and
a width of $39$~MeV for the $D_{13}(1700)$, which are very close to the PDG 
limits. Uncertainties of $21$~MeV for the mass and $12$~MeV for the width were 
estimated by examining the variation in the $3/2^-$ parameters while using 
various parameter values (all within the PDG limits) for the $5/2^+$, 
along with fitting different sub-ranges in $W$ of the phase motion.

%
%

A single channel analysis is not the best environment for
extracting precise resonance parameters; however, the qualitative agreement of
the phase motion obtained from the PWE's to that of the two PDG states is 
very suggestive of their presence in our data.
Recall that the phase parameters were each started pseudo-randomly in the range
$[0,2\pi)$ in each $W$ bin.
Yet the results are in good qualitative agreement with the phase motion 
expected using simple constant width Breit-Wigner distributions for the PDG 
$F_{15}(1680)$ and $D_{13}(1700)$ states.

\subsubsection{\label{section:results:threshold:helicity}Production helicity 
  amplitudes}

We can also compare the production helicity couplings extracted from our
fits to those quoted by the PDG. The ratio of the helicity amplitudes is 
obtained using the production couplings, $g_{LS_i}^{J^P}$ in 
Eq.~(\ref{eq:resonance}), extracted by the fits and the $s$-channel production 
amplitudes. Due to the nature of the covariant formalism, these amplitudes
are energy-dependent; {\em i.e.} the ratio of the helicity amplitudes is 
a function of $W$. The PDG quotes these values at the resonance masses.
For the $D_{13}(1700)$, the PDG reports the ratio of the helicity amplitudes 
as~\cite{pdg}
\begin{equation}
  \frac{A_{3/2}}{A_{1/2}} = 0.11 \pm 1.34.
\end{equation}
Our fits extract this value to be in the range $[-0.06,0.13]$, depending on
the mass of the $J^P = 3/2^-$ state. This is in a very good agreement with the
PDG value.

The ratio of the helicity amplitudes for the $F_{15}(1680)$ extracted by our 
fits is consistent with the PDG~\cite{pdg} value. However, projecting this ratio from 
the $p\omega$ threshold down to the required mass makes obtaining a precise
quantitative value difficult.

\subsubsection{\label{section:results:threshold:compare}
  Comparison to observables}

Fit~I consists of three production mechanisms:
(1)~OTL $t$-channel terms (with no free parameters), 
which are dominated by $\pi^0$ exchange in this energy range;
(2)~$J^P = 3/2^-$ $s$-channel waves, whose extracted parameters are consistent
with the PDG $D_{13}(1700)$ state;
(3)~${J^P = 5/2^+}$ $s$-channel waves, whose extracted parameters are 
consistent with the PDG $F_{15}(1680)$ state (at least, near threshold).
This is almost certainly not all of the physics contributing to 
$\omega$ photoproduction in this energy range. 
Thus, we would not expect Fit~I to provide a perfect description of our data. 
Before we examine the quality of Fit~I, we note that the OTL $t$-channel terms 
do provide a good description of our forward data in the energy regime of 
Fit~I.

\begin{figure*}[h!]\centering
\includegraphics[width=0.99\textwidth]{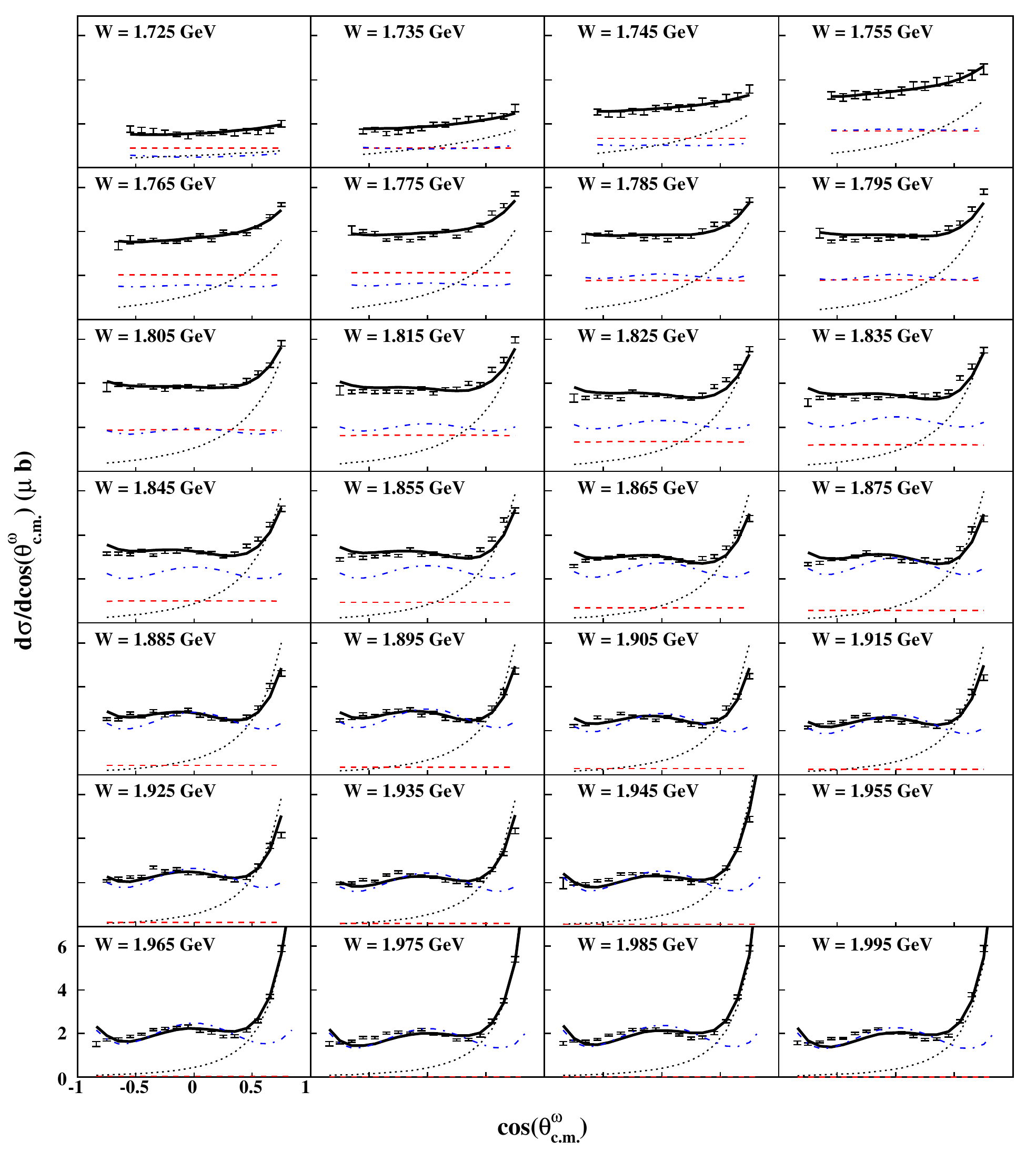}
\caption[]{
  (Color Online)
  $d\sigma/d\cos{\theta^{\omega}_{c.m.}} (\mu b)$ 
  vs $\cos{\theta^{\omega}_{c.m.}}$: 
  PWA results from Fit I (solid black line), compared to our 
  measurements~\cite{williams-prc}.
  The individual contributions from the $J^P = 3/2^-$ wave (dashed red line),
  $J^P = 5/2^+$ wave (dashed-dotted blue line) 
  and OTL $t$-channel terms (dotted black line) are also shown.
  The lack of data reported in the $W=1.955$~GeV bin is due to
  normalization issues~\cite{williams-prc}.
}
\label{fig:fit1:compare-dsigma}
\end{figure*}

\begin{figure*}[h!]\centering
\includegraphics[width=0.99\textwidth]{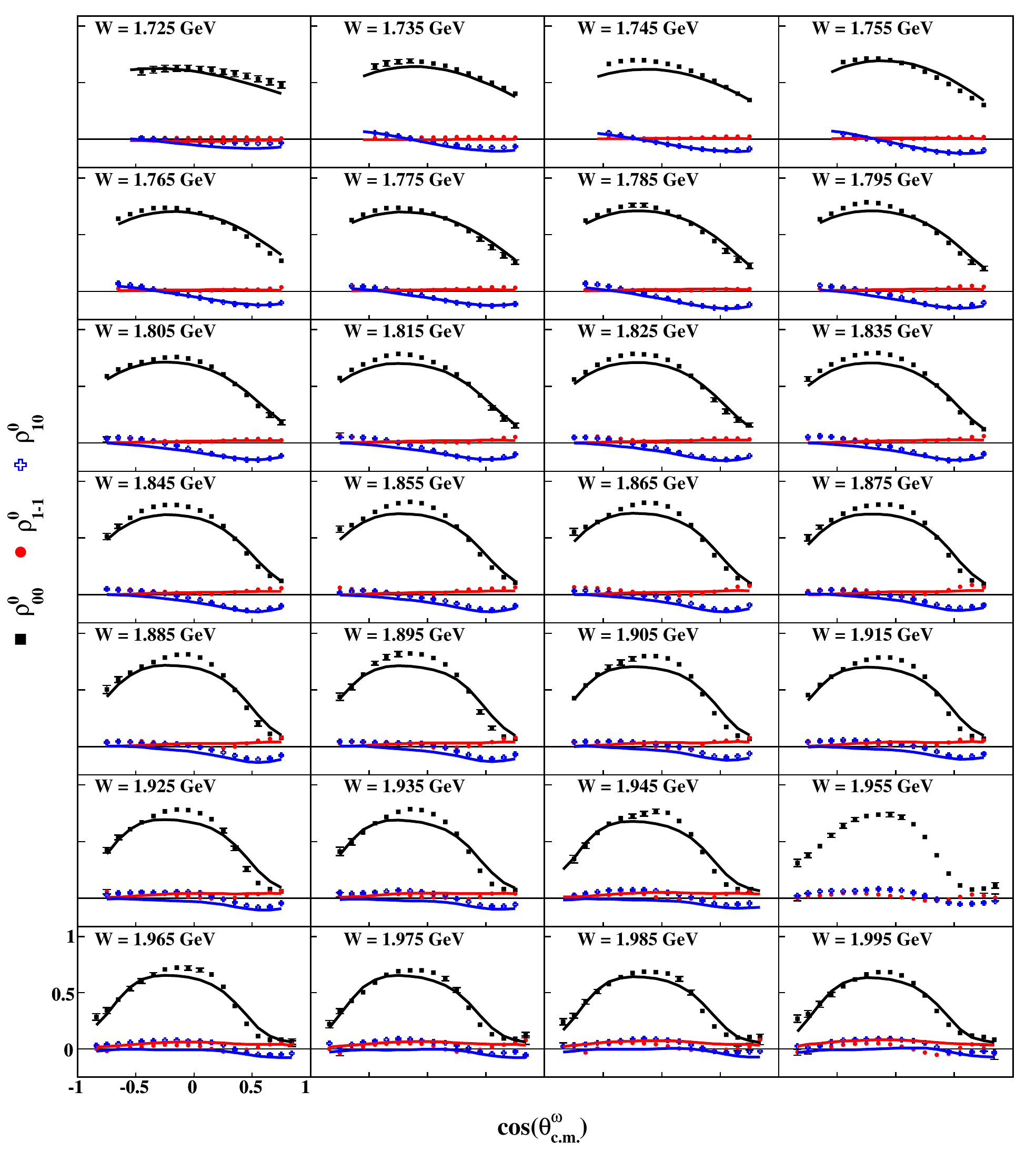}
\caption[]{
  (Color Online)
  $\rho^0_{MM'}$ 
  vs $\cos{\theta^{\omega}_{c.m.}}$: 
  PWA results from Fit I: $\rho^0_{00}$ (solid black line), 
  $\rho^0_{1-1}$ (solid red line), $Re\rho^0_{10}$ (solid blue line), 
  compared to our 
  measurements~\cite{williams-prc}.
 The lack of a fit in the $W=1.955$~GeV bin is due to
  normalization issues~\cite{williams-prc}.}  
\label{fig:fit1:compare-rho}
\end{figure*}

Figures~\ref{fig:fit1:compare-dsigma} and \ref{fig:fit1:compare-rho} show 
comparisons of the differential cross sections and spin density matrix elements
extracted from the PWA fits compared to our measurements~\cite{williams-prc}. 
Recall that we do not fit to the experimental observables directly. 
We perform event-based fits to the data used to obtain the measured results.
The forward cross section and polarization observables are very well described
in this energy range, confirming that $\pi^0$ exchange in the $t$-channel
does dominate the amplitude in this region.
There is a discrepancy in the description of the cross section at backwards 
angles that increases with energy.
This could be due to the lack of $u$-channel terms. It could also be a 
signature of unaccounted-for $s$-channel amplitudes. 

Though we did not start off by including known PDG resonance states, the fit 
has extracted evidence for them from our data.  We also note here that the 
large-angle cross section at $W = 1.8$~GeV is virtually flat.  Without polarization 
information, the production mechanism could have easily been mistaken for a 
$J = 1/2$ wave. This demonstrates the importance of the spin density matrix 
elements.

The quality of the description of the observables decreases slightly with
increasing energy. This signifies that there is another production 
mechanism that is not accounted for in the fit. This is expected due to the
limited number of waves included in the PWA.
Adding additional waves improves the description of the data but has 
virtually no effect on the strengths and phase motion of the two $s$-channel 
waves presented in this section. Thus, the conclusions drawn about resonance 
contributions are robust, and do not change when additions are made to the wave set.  
See Section~\ref{section:results:additional-states:fitiii} for results obtained
by adding an additional $s$-channel wave to this fit.

\subsection{\label{section:results:higher}Fit II: the higher mass region}

Our preliminary $s$-channel scans showed that the best fit using
two $s$-channel waves and the OTL $t$-channel terms in the energy range 
$2$~GeV${<W<}\, 2.4$~GeV is 
obtained using waves with $J^P = 5/2^+,7/2^-$. 
To extract any possible resonance contributions in these waves, the same 
procedure used for Fit~I was employed.

\subsubsection{\label{section:results:higher:cs_and_dphase}Cross sections
  and phase motion}

\begin{figure}[h!]
\begin{center}
  \includegraphics[width=0.5\textwidth]{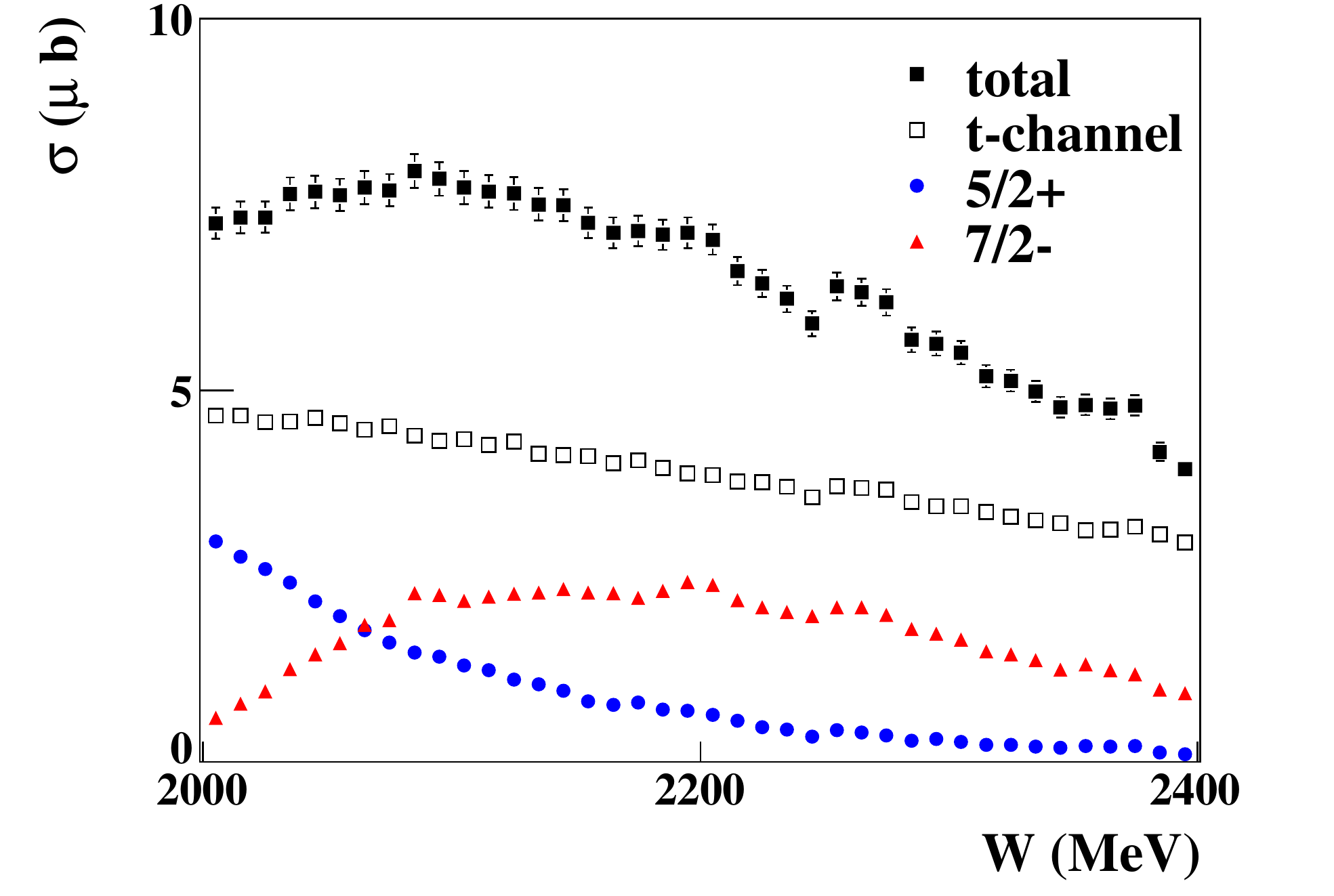}
\caption[]{
  Results from Fit II: $\sigma (\mu b)$ vs $W$(MeV): Total cross sections from 
  all of
  the waves in the fit (filled squares), only $t$-channel waves (open squares),
  only $J^P = 5/2^+$ waves (circles) and only $J^P = 7/2^-$ waves (triangles).
  The cross section extracted for $J^P = 5/2^+$ is consistent with the tail of
  a lower mass state (as seen in Fit~I). The $J^P = 7/2^-$ cross section is 
  indicative of a state near $2.2$~GeV. The errors are purely statistical.
}
\label{fig:fit2:sigma}
\end{center}
\end{figure}

As in Fit~I, the strength and phase of each $s$-channel wave were completely 
free to vary in each energy bin in the PWE's. 
The cross sections extracted for the $s$-channel waves, shown in
Fig.~\ref{fig:fit2:sigma}, are consistent with the tail of a ${J^P = 5/2^+}$
state with a mass below 2~GeV (as seen in Fit~I) and a 
${J^P = 7/2^-}$ state with a mass near 2.2~GeV.
The PDG lists a state consistent with this hypothesis: the
4-star~$G_{17}(2190)$, which currently is only listed as having 1-star 
coupling to $\gamma p$.

\begin{figure}[h!]
\begin{center}
  \includegraphics[width=0.5\textwidth]{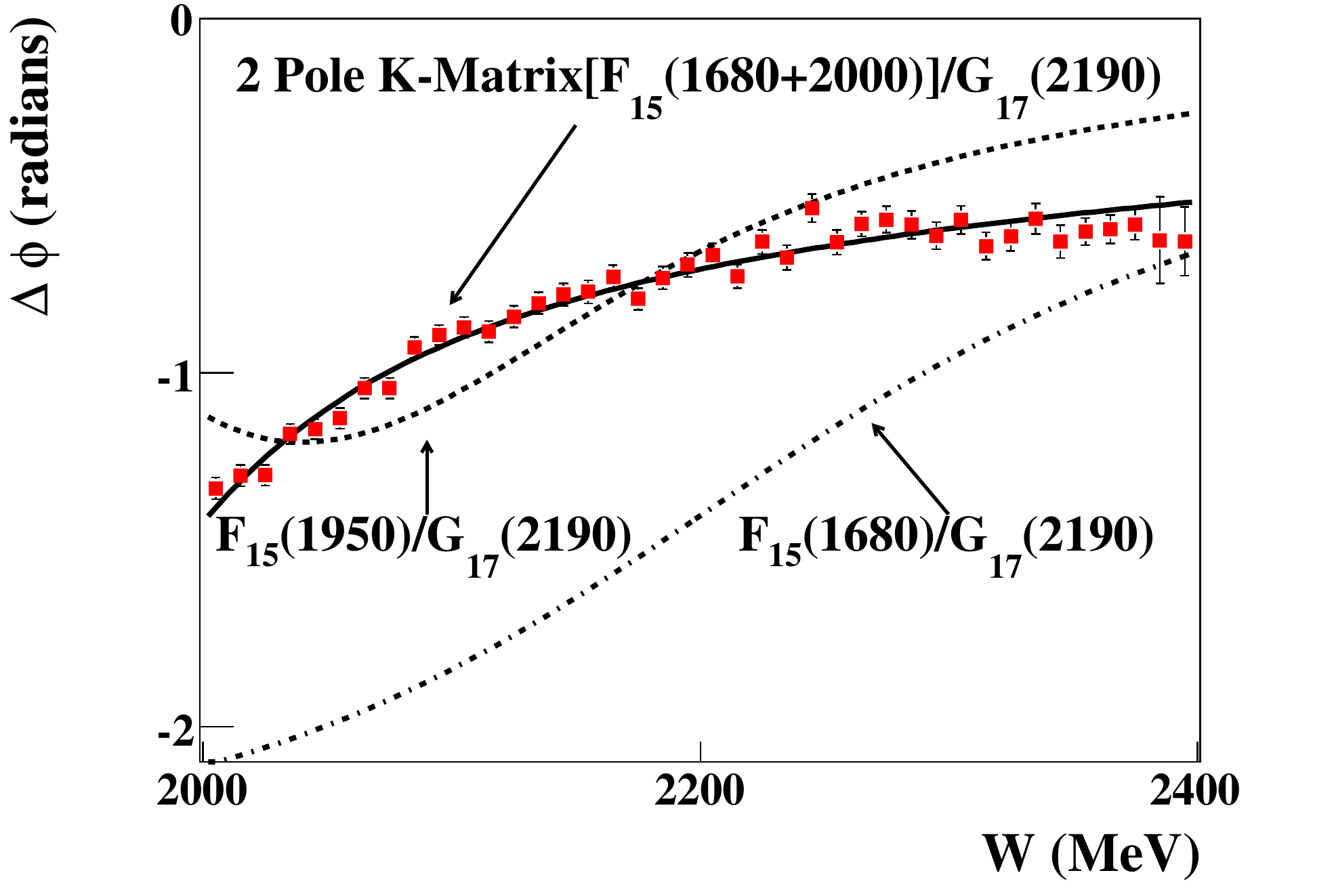}
\caption[]{
  Results from Fit II: $\Delta\phi = \phi_{7/2^-} - \phi_{5/2^+}$(radians) vs 
  $W$(MeV): 
  The dot-dashed line is the phase motion expected using constant 
  width Breit-Wigner distributions and the parameters quoted by the PDG for the 
  $F_{15}(1680)$ and $G_{17}(2190)$.
  The dashed line required the $J^P=7/2^-$ parameters to be within the PDG 
  limits for the $G_{17}(2190)$, while allowing the $J^P=5/2^+$ parameters
  to vary freely. The solid line used a constant width Breit-Wigner distribution for the
  $G_{17}(2190)$, but a 2-pole single channel K-matrix for the 
  $J^P=5/2^+$ wave. The parameters obtained from these fits are listed in
  the text. The error bars are purely statistical.
}
\label{fig:fit2:dphase}
\end{center}
\end{figure}

Figure~\ref{fig:fit2:dphase} shows the phase motion between the two
$s$-channel waves extracted from the PWE's. 
Our results do not agree with those expected from the PDG 
$F_{15}(1680)$ and $G_{17}(2190)$ states, assuming their mass dependencies are 
well described by the constant width Breit-Wigner line shape described in
Eq.~(\ref{eq:bw}). 
Allowing the $J^P = 5/2^+$ parameters to vary freely gives us better 
agreement and yields a mass around 1.95~GeV.
The presence of a second state in the $J^P = 5/2^+$ wave near this mass would
have virtually no effect on the results obtained in Fit~I; however, it would 
mean that the use of a Breit-Wigner distribution in the energy range where both $5/2^+$
states are contributing significantly, {\em e.g.} the energy region examined
in Fit~II, is invalid. 

Instead, we can employ a two-pole single channel K-matrix for the $5/2^+$
states of the form~\cite{k-matrix}
\begin{equation}
  \label{eq:k-matrix}
  K(s) = \sum\limits_{\alpha = 1}^{2}
  \frac{g_{\alpha p\omega}^2 B^2_{\ell}(s)}{w^2_{\alpha} - s},
\end{equation}
where $w_{\alpha}$ and $g_{\alpha p\omega}$ are the K-matrix 
resonance masses and 
coupling constants to the $p\omega$ final state and $B_{\ell}$ are the 
centrifugal barrier factors (see, {\em e.g.},~\cite{anisovich}). 
The mass dependence of the
amplitude is then written in terms of the production vector, $P$, and 2-body
phase-space factor, $\rho$, as
\begin{equation}
  \mathcal{R}_{5/2^+}(s) = P\left(1 - i \rho K\right)^{-1},
\end{equation}
where 
\begin{equation}
  P = \sum\limits_{\alpha = 1}^{2}
  \frac{g_{\alpha p\gamma}g_{\alpha p\omega} B_{\ell}(s)}{w^2_{\alpha} - s},
\end{equation}
with production coupling constants $g_{\alpha p\gamma}$ and
\begin{equation}
  \rho = \frac{\sqrt{(s - (w_{\omega} + w_p)^2)(s - (w_{\omega} - w_p)^2)}}{s}.
\end{equation}

For this MDF, we required the  $J^P = 7/2^-$ parameters to be within the
limits quoted by the PDG for the $G_{17}(2190)$. One of the $J^P = 5/2^+$
K-matrix poles was required to be consistent with the $F_{15}(1680)$.  
The exact location of this pole depends on how one treats the opening of the
$p\omega$ threshold (this is a single channel analysis). The parameters of the
second $J^P = 5/2^+$ K-matrix pole were obtained from the MDF yielding
$1930$~MeV for the mass and $100$~MeV for the width. 
Even with these constraints,
the results provide a very good description of the phase motion obtained from
the PWE. 

In principle, poles in the T-matrix and poles in the K-matrix can be quite
different. The relationship between the two can also depend on the
specific K-matrix model employed. For these reasons, care must be taken when
interpreting the parameters obtained for the ``second'' $5/2^+$ resonance. 
While the K-matrix parameters may not coincide exactly 
with the physical T-matrix values, the observed strength in this wave and its
phase motion relative to the $G_{17}(2190)$ support it having a mass 
around $1.9$--$2$~GeV and a width of approximately $200$--$300$~MeV.  
These values are in good agreement 
with the missing  $F_{15}(2000)$ state predicted by~\cite{capstick}.
A check using the (unitarity violating) two Breit-Wigner prescription also 
resulted in $5/2^+$ resonance parameters in this range. To extract precise
resonance parameters for this state (and to confirm its existence), a 
coupled-channel analysis should be employed.

%
%

\subsubsection{\label{section:results:higher:helicity}Production helicity 
  amplitudes}

The ratio of the helicity amplitudes for the $F_{15}(1680)$ was discussed
in Section~\ref{section:results:threshold:helicity}. Without employing a model
we cannot separate out the possible missing $F_{15}(2000)$ production
amplitudes. 
The ratio of the helicity amplitudes for the $G_{17}(2190)$ is extracted to
be
\begin{equation}
  \frac{A_{3/2}}{A_{1/2}} = -0.17 \pm 0.15.
\end{equation}
Due to its 1-star coupling to $\gamma p$, the PDG does not quote a value for 
this ratio.

\subsubsection{\label{section:results:higher:compare}
  Comparison to observables}

As in Fit~I, we do not expect the limited number of waves used in Fit~II
to include all of the physics at these energies. Thus, we again do not 
expect to provide a perfect description of the observables in this energy 
regime. Before we examine Fit~II, we note that the OTL $t$-channel terms 
provide a good description of our forward cross sections; however, there are
some noticeable discrepancies with the spin density matrix elements at these
energies~\cite{williams-prl}.

\begin{figure*}[h!]
\begin{center}
  \includegraphics[width=0.99\textwidth]{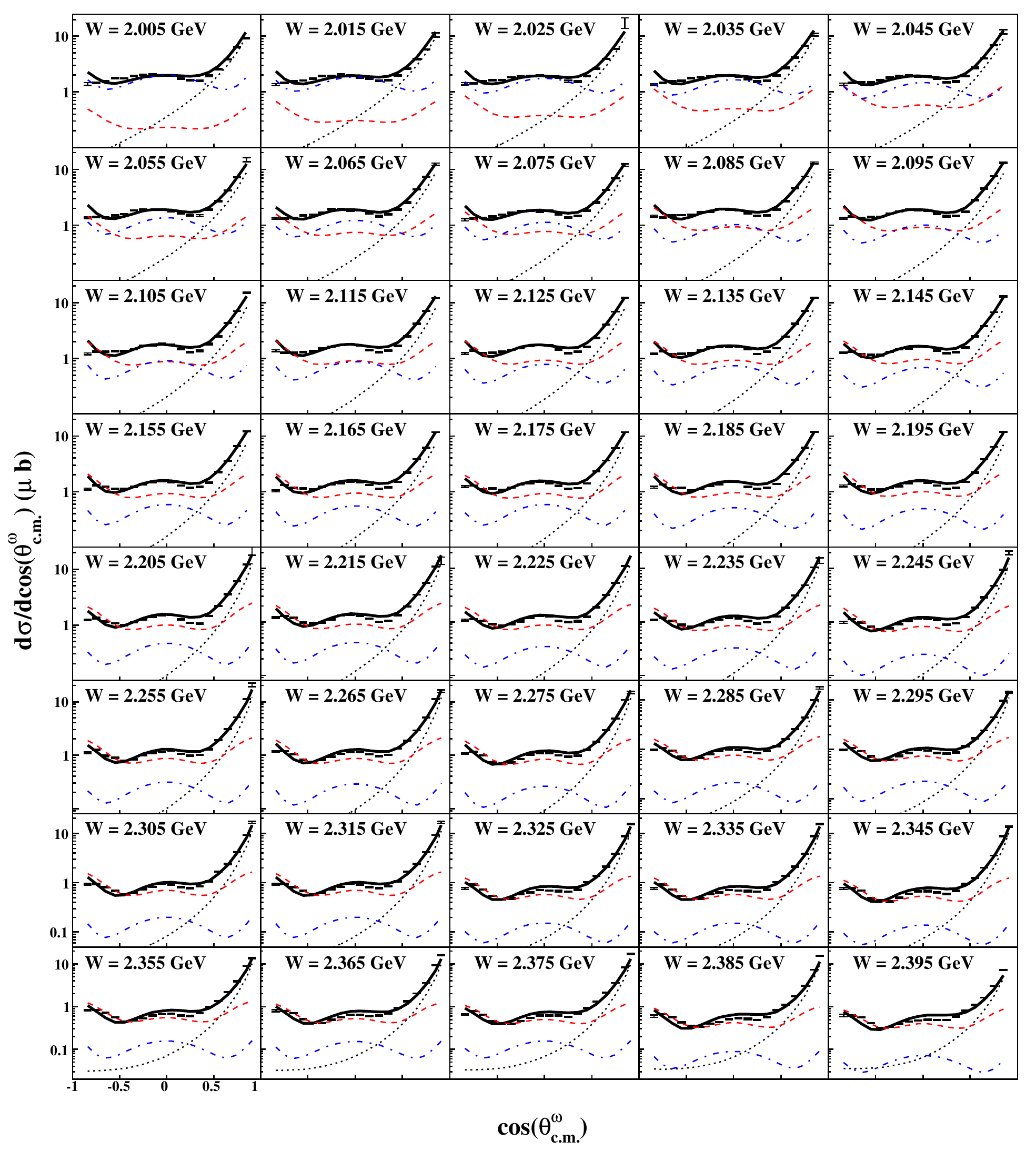}
\caption[]{
  (Color Online)
  $d\sigma/d\cos{\theta^{\omega}_{c.m.}} (\mu b)$ 
  vs $\cos{\theta^{\omega}_{c.m.}}$: 
  PWA results from Fit II (solid black line), compared to our 
  measurements~\cite{williams-prc}.
  The individual contributions from the $J^P = 7/2^-$ wave (dashed red line),
  $J^P = 5/2^+$ wave (dashed-dotted blue line) 
  and OTL $t$-channel terms (dotted black line) are also shown.
}
\label{fig:fit2:compare-dsigma}
\end{center}
\end{figure*}

\begin{figure*}[h!]
\begin{center}
  \includegraphics[width=0.99\textwidth]{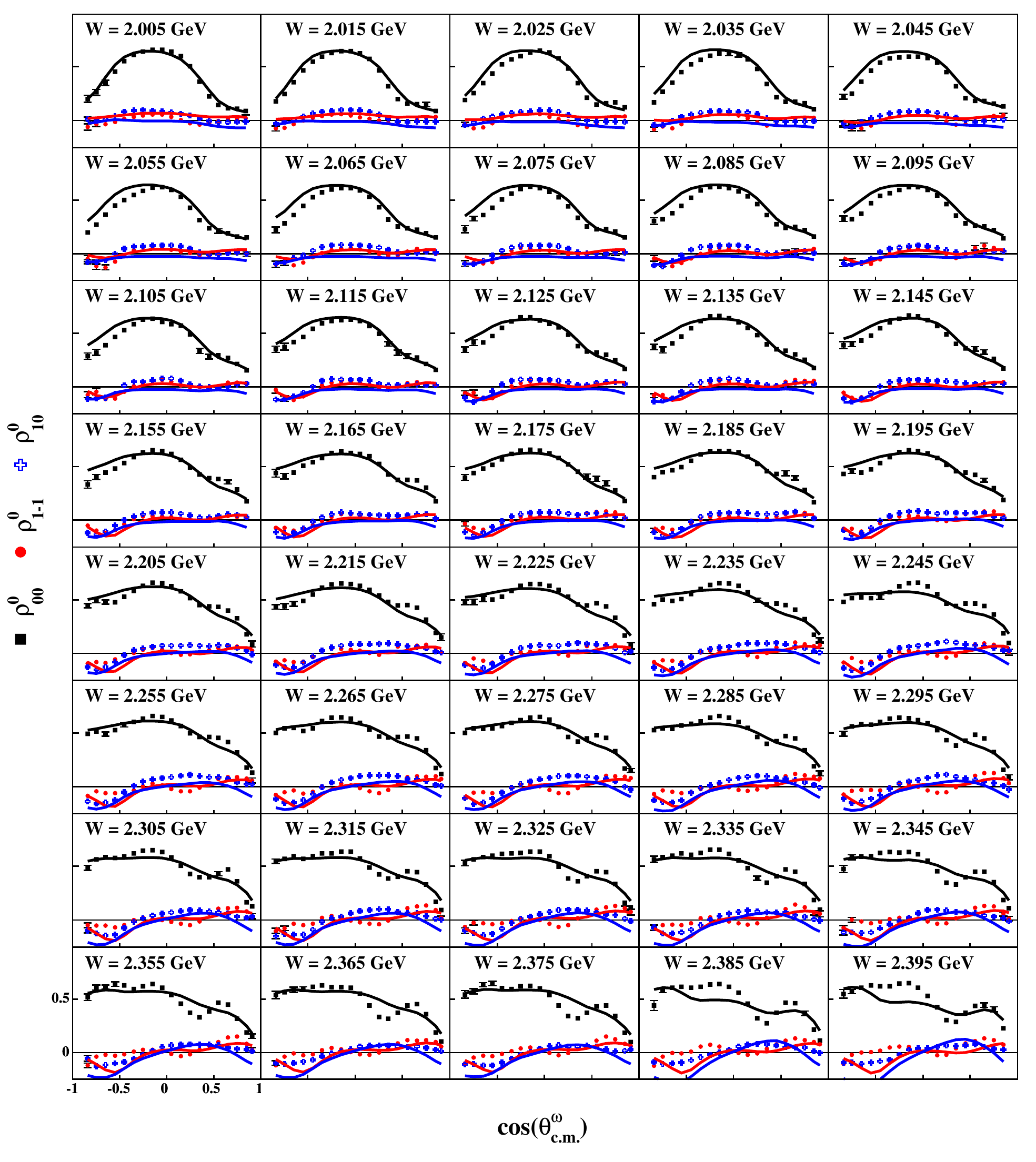}
\caption[]{
  (Color Online)
  $\rho^0_{MM'}$ 
  vs $\cos{\theta^{\omega}_{c.m.}}$: 
  PWA results from Fit II: $\rho^0_{00}$ (solid black line), 
  $\rho^0_{1-1}$ (solid red line), $Re\rho^0_{10}$ (solid blue line), 
  compared to our 
  measurements~\cite{williams-prc}. See text for discussion.
}
\label{fig:fit2:compare-rho}
\end{center}
\end{figure*}

Figures~\ref{fig:fit2:compare-dsigma} and \ref{fig:fit2:compare-rho} show 
the differential cross sections and spin density matrix elements
extracted from the PWA fits compared to our 
measurements~\cite{williams-prc}.
As the energy increases, the quality of the descriptions of the spin density 
matrix elements decreases. 
This is partly due to the issues with the non-resonant model employed
(as discussed in Section~\ref{section:amps:non-resonant}).
At these energies, the OTL $t$-channel terms begin to fail to adequately 
describe the polarization observables.
The discrepancies in the backwards direction could be due to the
lack of inclusion of any $u$-channel terms.
The effects on our results, {\em i.e.} the effects on conclusions drawn about
resonance contributions, due to possible issues with the non-resonant terms
are discussed in Section~\ref{section:sys}.

The lack of perfect description of the data signifies that there are other 
production mechanisms that are not accounted for in the fit. 
This is, again, expected due to the limited number of waves included in the 
PWA. 
Adding additional waves improves the description of the data.  The 
strengths and phase motion of the two $s$-channel waves presented in this 
section become noisier in the presence of these additional waves; however, the 
conclusions drawn about resonance contributions are unaffected by additions to 
the wave set.  
We are unable to determine from our PWE's which additional waves may coincide 
with
unaccounted-for physical processes; thus, we do not present them here.
Perhaps future measurement of additional polarization observables in
this energy range might help determine the nature of these waves.

\subsection{\label{section:results:additional-states}Evidence for additional
  resonance states}

One of the prime motivating factors in undertaking this study was to search
for missing resonances. The strongest evidence for resonance 
contributions to $\omega$ photoproduction found in Fits~I and II was for 
well-known PDG states. Suggestive evidence was also found for a 
missing $F_{15}(2000)$ state. Below we examine possible additional resonance
contributions.

\subsubsection{\label{section:results:additional-states:fitiii}
  Fit III: the $3/2^+$ wave}

\begin{figure*}[h!]
\begin{center}
\subfigure[]{ 
  \includegraphics[width=0.49\textwidth]{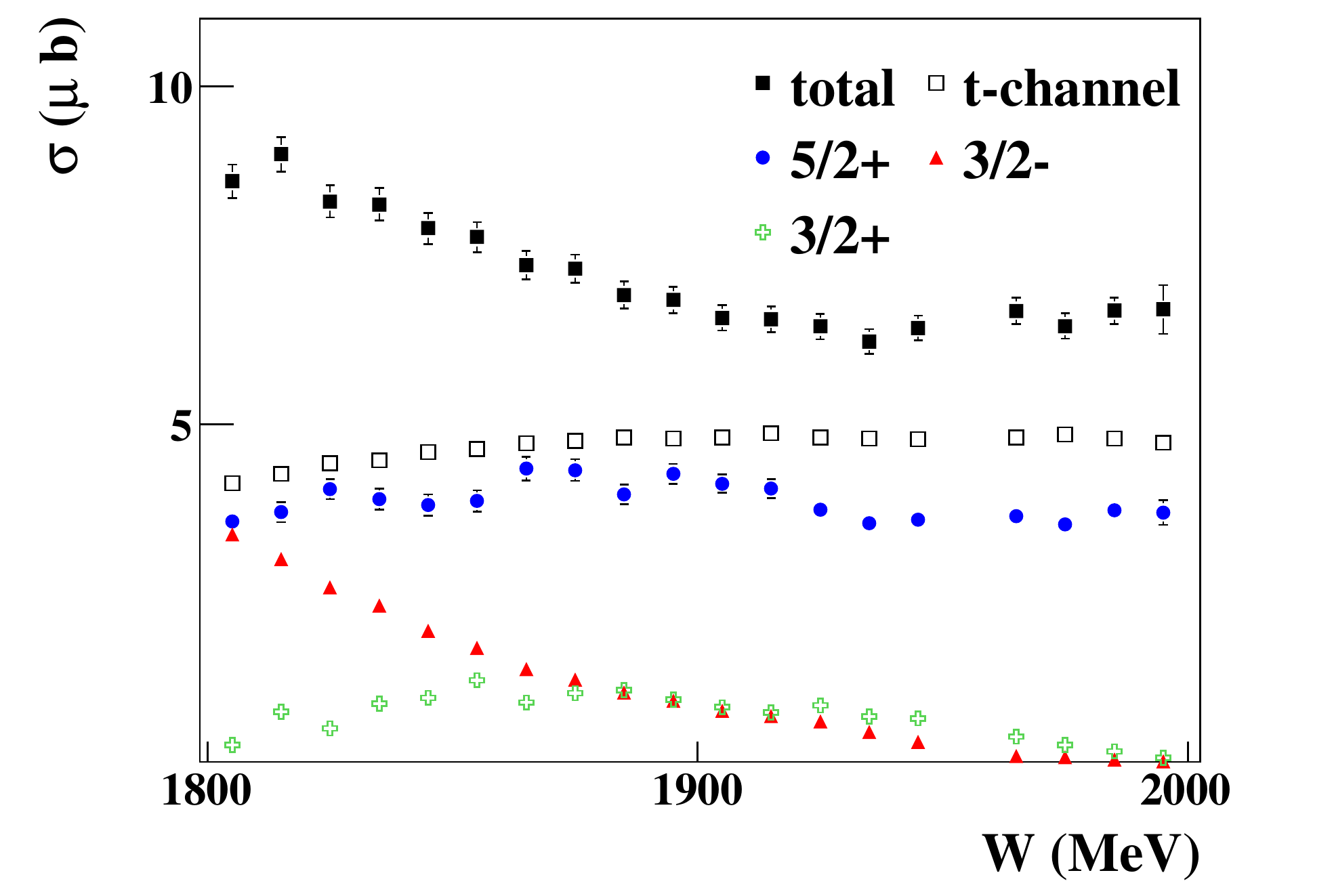}
}
\hspace{-0.05\textwidth}
\subfigure[]{ 
  \includegraphics[width=0.49\textwidth]{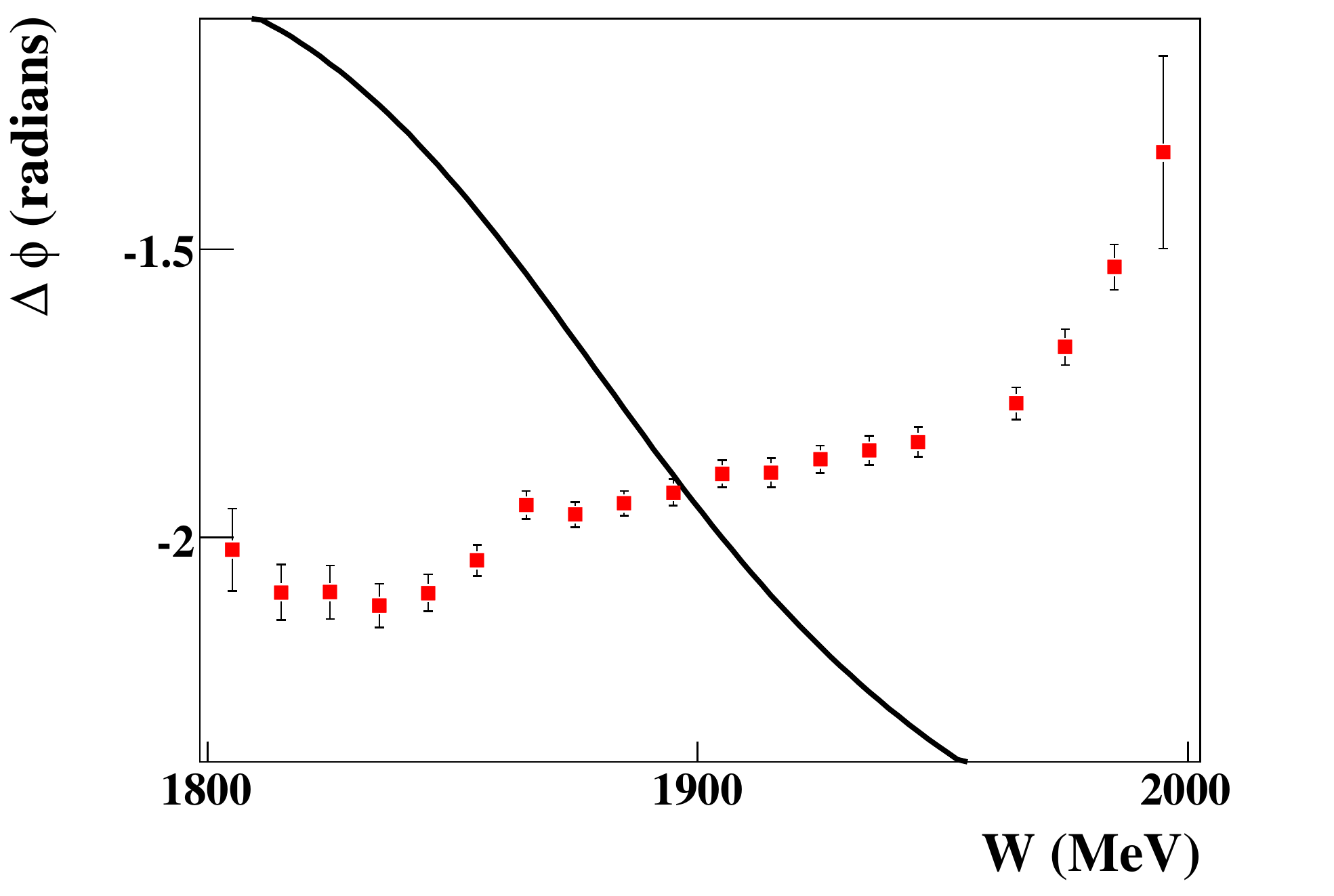}
} 
\subfigure[]{ 
  \includegraphics[width=0.49\textwidth]{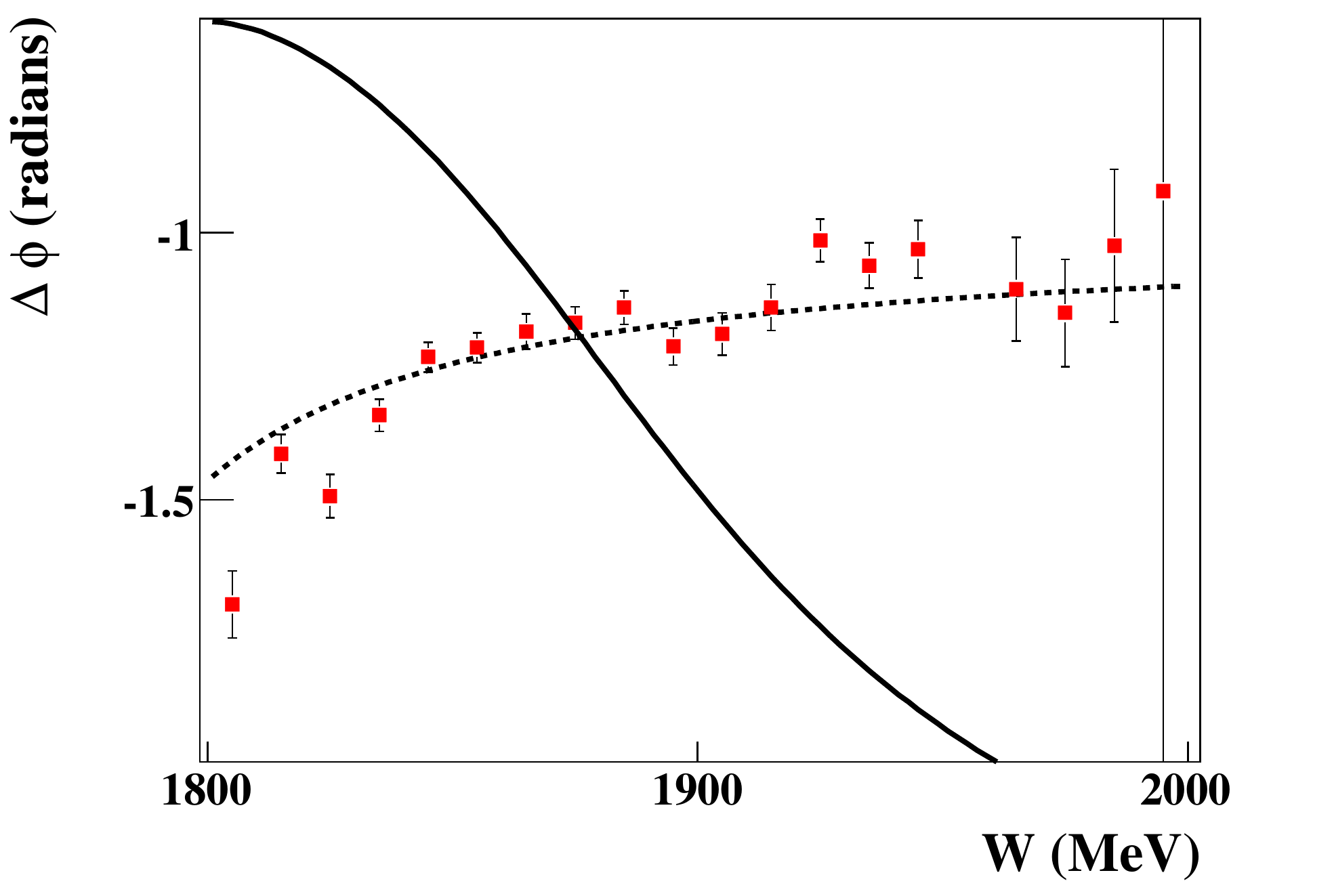}
} 
\caption[]{
  Results from Fit III:
  (a) $\sigma(\mu b)$ vs $W$(MeV): Total cross sections extracted from all of 
  the waves in the
  fit (filled black squares), only $t$-channel waves (open squares), 
  only $5/2^+$ waves (circles), 
  only $3/2^-$ waves (triangles) and only $3/2^+$ waves (crosses). 
  (b) $\Delta\phi = \phi_{5/2^+} - \phi_{3/2^+}$(radians)~vs~$W$(MeV).
  (c)~$\Delta\phi = \phi_{3/2^-} - \phi_{3/2^+}$(radians) vs $W$(MeV).
  The solid curves show the phase motion expected assuming the 
  $J^P = 3/2^+$ has Breit-Wigner parameters $M_{3/2^+} = 1875$~MeV and 
  $\Gamma_{3/2^+} = 150$~MeV while locking the $J^P = 3/2^-$ and $J^P = 5/2^+$
  parameters to be those of the $D_{13}(1700)$ and $F_{15}(1680)$, 
  respectively.
  The phase motion obtained for the $3/2^+$ is not consistent with a single
  resonant state.
  The dashed line on (c) represents using the $D_{13}(1700)$ Breit-Wigner 
  parameters for the $3/2^-$ and a single channel two-pole K-matrix for the
  $3/2^+$. There is enough freedom to describe the data 
  (see text for discussion).
  All error bars are purely statistical.
}
\label{fig:fit3:results}
\end{center}
\end{figure*}

Quark model calculations predict three missing resonances with $J^P = 3/2^+$
in the energy range of Fit~I which couple to 
$p\omega$~\cite{capstick}.
Figure \ref{fig:fit3:results} shows the cross sections and phase motion
obtained if we add a $J^P = 3/2^+$ wave to the PWE in Fit~I. 
Below $1800$~MeV, the range of production angles over which the CLAS has 
acceptance is limited. This makes it difficult to cleanly separate 
contributions from three waves; thus, this energy range has been excluded from
these fits. 
The strengths and phases of the $J^P = 3/2^-,5/2^+$ waves are virtually
unaltered by the addition of the extra $s$-channel wave.
The cross section of the $J^P = 3/2^+$ wave does show some fairly smooth 
structure; however, its phase motion, relative to the other two resonant 
states, is not consistent with a single constant width Breit-Wigner hypothesis.
If we instead perform a MDF using the K-matrix formalism described in 
Section~\ref{section:results:higher:cs_and_dphase} for the $J^P = 3/2^+$ wave,
the phase motion between the $J^P = 3/2^-,3/2^+$ waves is well described.
The poles in the K-matrix are at 1850~MeV and 1950~MeV.
We could apply the same procedure for the  $J^P = 5/2^+,3/2^+$ phase motion;
however, both waves would have K-matrices and the number of free parameters
would leave the fit under-constrained. 

\begin{figure*}[h!]
\begin{center}
  \includegraphics[width=0.99\textwidth]{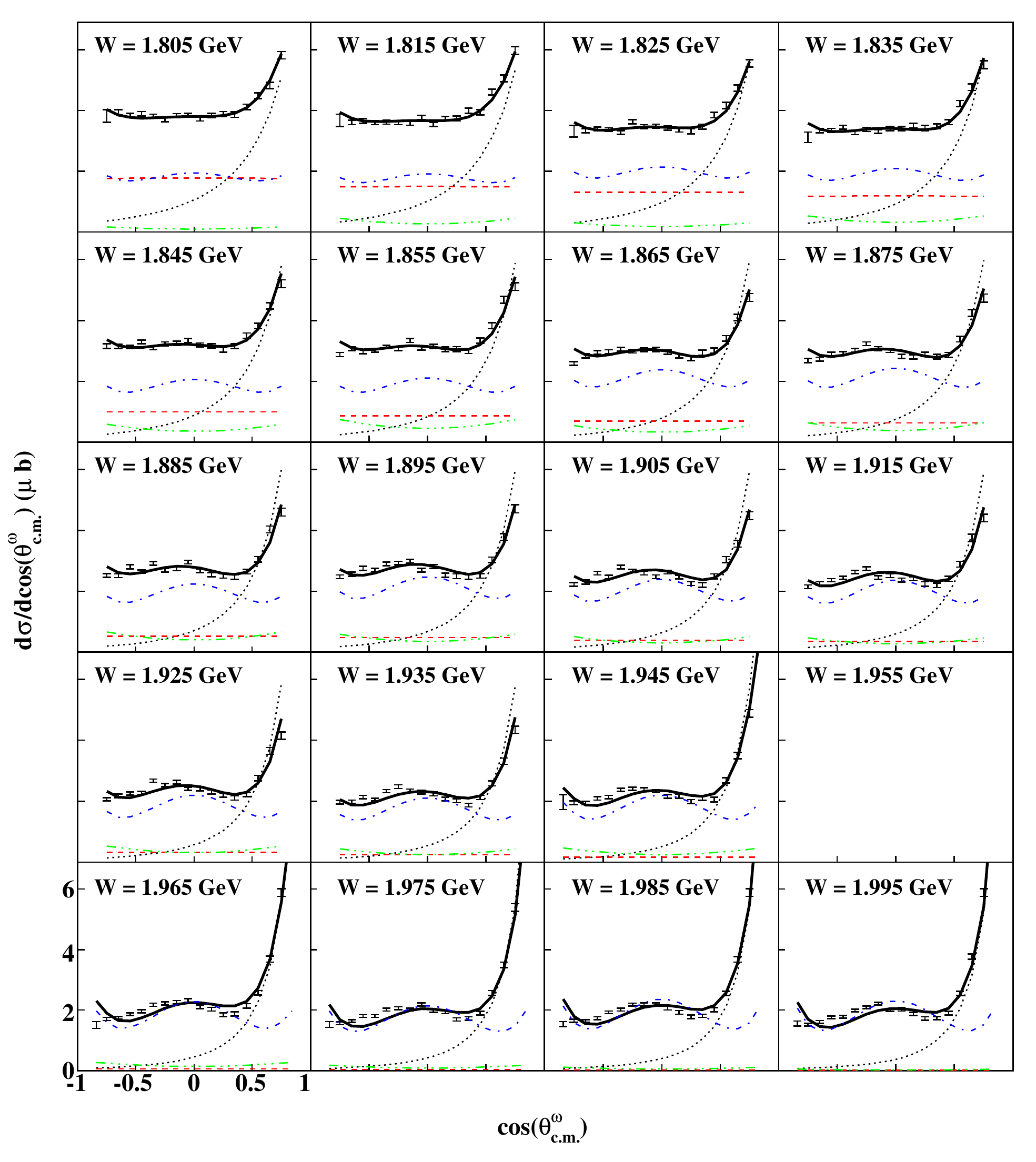}
\caption[]{
  (Color Online)
  $d\sigma/d\cos{\theta^{\omega}_{c.m.}} (\mu b)$ 
  vs $\cos{\theta^{\omega}_{c.m.}}$: 
  PWA results from Fit III (solid black line), compared to our 
  measurements~\cite{williams-prc}.
  The individual contributions from the $J^P = 3/2^-$ wave (dashed red line),
  $J^P = 5/2^+$ wave (dashed-dotted blue line), 
  the $J^P = 3/2^+$ wave (dashed-triple-dotted green line)
  and OTL $t$-channel terms (dotted black line) are also shown.
  The lack of data reported in the $W=1.955$~GeV bin is due to
  normalization issues~\cite{williams-prc}.
}
\label{fig:fit3:compare-dsigma}
\end{center}
\end{figure*}

\begin{figure*}[p]
\begin{center}
  \includegraphics[width=0.99\textwidth]{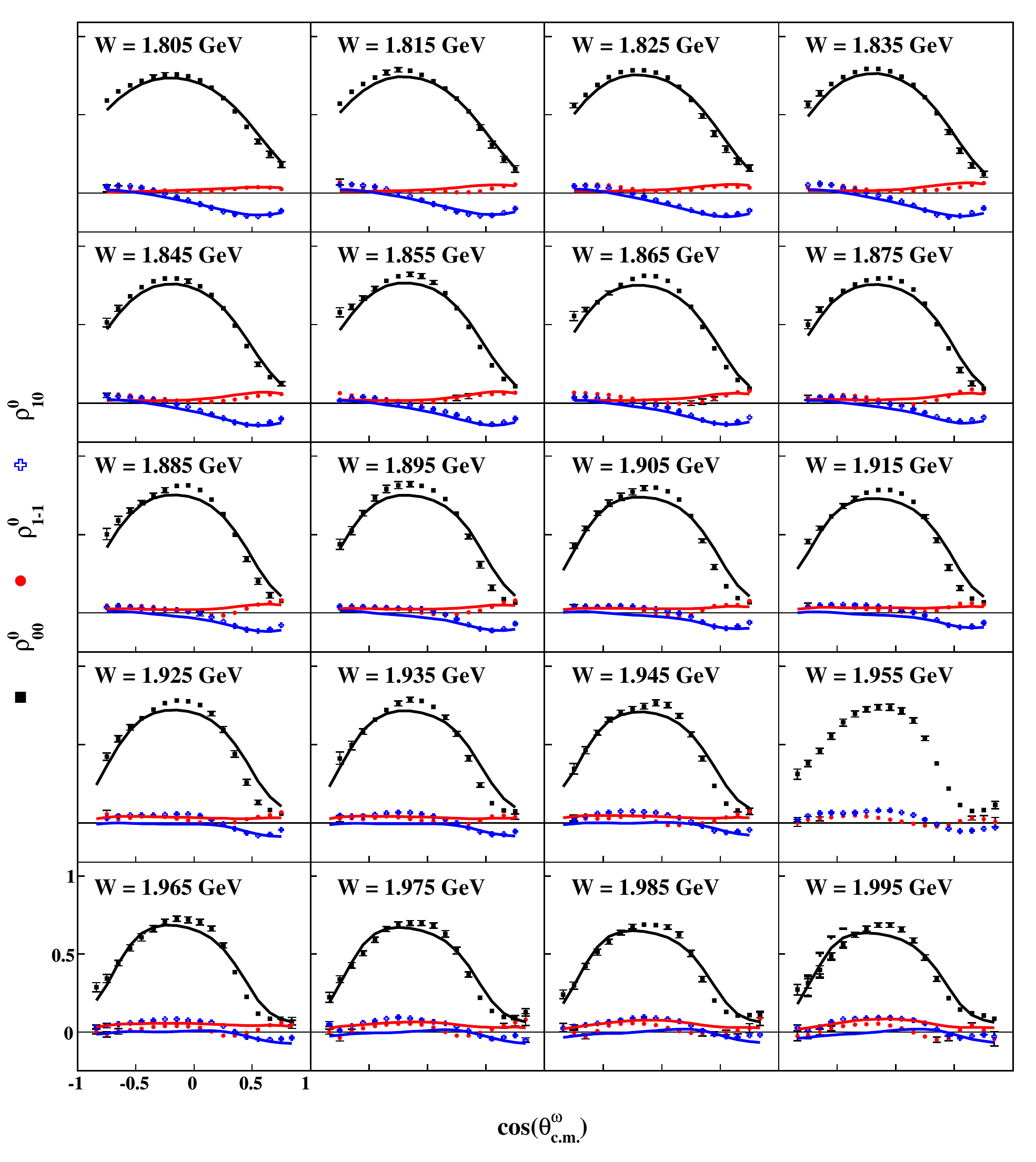}
\caption[]{
  (Color Online)
  $\rho^0_{MM'}$ 
  vs $\cos{\theta^{\omega}_{c.m.}}$: 
  PWA results from Fit III: $\rho^0_{00}$ (solid black line), 
  $\rho^0_{1-1}$ (solid red line), $Re\rho^0_{10}$ (solid blue line), 
  compared to our 
  measurements~\cite{williams-prc}.
  No fit was performed in the $W=1.955$~GeV bin, 
  see text for details.
}
\label{fig:fit3:compare-rho}
\end{center}
\end{figure*}

Figures~\ref{fig:fit3:compare-dsigma} and \ref{fig:fit3:compare-rho} show the
comparisons of the PWA results with our measurements~\cite{williams-prc}.
The additional $3/2^+$ wave improves the description of the spin
density matrix elements obtained in Fit~I (see Fig.~\ref{fig:fit1:compare-rho}).
It is difficult to make firm conclusions about the ${J^P = 3/2^+}$ wave.
The observed strength suggests there is significant overlap of the scattering
amplitude with this partial wave; however, this is not sufficient evidence 
to claim resonance contributions. The phase motion of the $3/2^+$ wave relative
to the $3/2^-$ and $5/2+$ waves is not consistent with a single resonant state.
This does not, however, rule out the existence of the multiple states predicted
by the quark model. 

\subsubsection{\label{section:results:additional-states:others}
  Limitations of the mass-independent technique}

Numerous other fits that we have performed yielded inconclusive evidence for 
states of various spin-parities.
These fits are very similar to  Fit~III. Generally, smooth structures are found
in the extracted cross sections; however, the phase motion is inconsistent with
a single resonant state. It is possible that a number of resonant states exist 
which couple relatively strongly to $p\omega$. 
It is also possible that the smooth cross sections are simply the result of 
overlap of various partial waves with unaccounted-for non-resonant terms. 
It would appear that we have reached the limits of what our 
technique can extract from our data. 

More polarization information may be required to cleanly extract additional resonances. 
Improved theoretical input for the non-resonant (non $s$-channel) terms may also be 
necessary. While our studies have shown that the strong $s$-channel signals extracted by
this analysis are not affected by the way that the non-resonant terms are modeled 
(see Sec.~\ref{section:sys}), this is almost certainly not the case for weaker signals.  
This is particularly true at higher energies, where the current theoretical models do a 
poor job of describing the new CLAS data. The amplitudes that are currently being 
generated by several groups from a  coupled-channel approach (see, 
{\em e.g.},~\cite{paris-09})  may well allow for the extraction of much weaker 
resonance signals from these data.

\section{\label{section:sys}Systematic Studies}

\subsection{$s$-channel scans}

In Section~\ref{section:results:scans}, we found that the single $s$-channel
waves with the best likelihoods were ${J^P = 3/2^-}$ for ${W < 1.85}$~GeV
and $J^P = 5/2^+$ for 1.85~GeV${<W< 2}$~GeV. We proceeded to add single
$s$-channel waves to these fits to determine which had the best likelihoods;
these were the basis for the wave sets used in Fits I and II. 
We can also examine the ``discarded'' wave sets and examine the $s$-channel
contributions as a systematic check on our results. 

In the $W< 2$~GeV energy range, the $s$-channel waves with $J^P = 3/2^-,5/2^+$,
used in Fit~I, had the best likelihood of all two $s$-channel wave 
combinations (when combined with the OTL $t$-channel waves). In this fit,
the contributions extracted for the two $s$-channel waves were approximately 
equal in size for $W < 1.85$~GeV.
For all other $[3/2^-,J^P]$ combinations, the extracted contribution for the 
$J^P = 3/2^-$ wave was the bigger of the two $s$-channel terms in this energy 
range. 
In the 1.85~GeV$<W<$2~GeV energy range, the $5/2^+$ wave had the
larger of the two $s$-channel contributions for every $J^P$ used for the other
$s$-channel wave.
Fits were also run using all two $s$-channel wave combinations (with the 
OTL $t$-channel terms) over the entire
energy range. The contributions of the waves presented in this paper were 
consistent, regardless of which other waves they were fit with.

The robustness of the results presented in this paper was also tested by 
performing the PWE's with larger wave sets.  The presence of any additional 
$J \leq 5/2$ wave does not effect the conclusions drawn about resonance 
contributions to Fit~I or Fit~II.  Fits run with very large wave sets that
included all $s$-channel waves with $J \leq 9/2$ also confirm the large
contributions from the $J^P = 3/2^-$ and $5/2^+$ below 2~GeV and from 
$J \geq 7/2$ waves around 2.2~GeV; however, with this many waves it was not
possible to unambiguously determine the spin-parity of the large $J$ 
contribution.

\subsection{Including $u$-channel terms}

Another possible cause of systematic effects is our lack of inclusion of any
$u$-channel terms.
For $W < 2$ GeV, any $u$-channel contribution must be small due to the 
lack of any visible peak in the backwards cross section. 
Thus, the conclusions drawn 
from Fit~I are independent of whether or not $u$-channel terms are included.
The same cannot be said for the energy regime of Fit~II. 
In~\cite{williams-prl}, 
we were able to modify the $u$-channel parameters of the 
Oh, Titov and Lee model to better describe our highest energy data; however, 
these amplitudes were not included in our PWA fits due to a lack of 
confidence in the assumptions used to obtain the parameters.

\begin{figure}[t]
\begin{center}
  \includegraphics[width=0.45\textwidth]{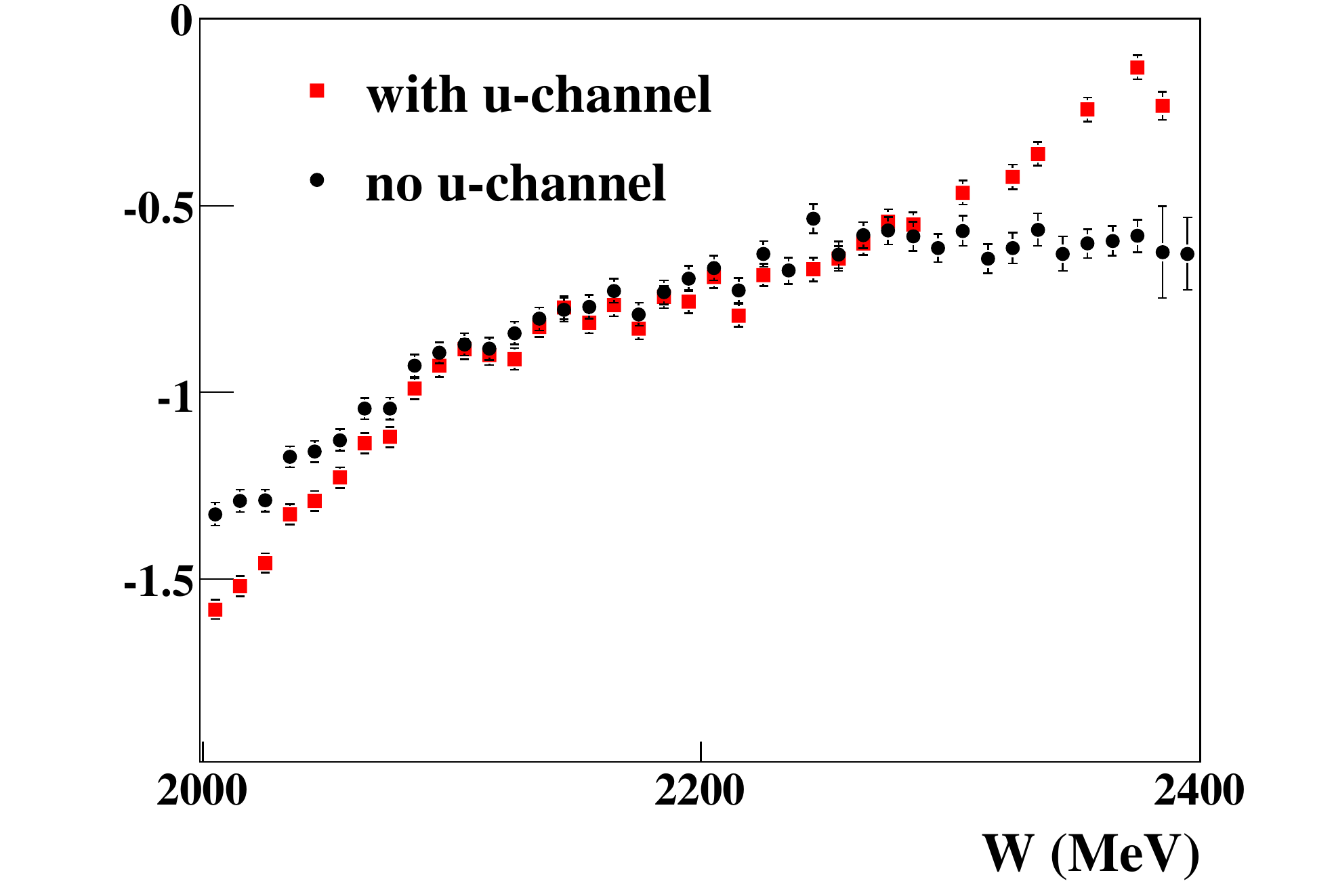}
\caption[]{
  Fit II $+u$-channel:
  $\Delta\phi = \phi_{7/2^-} - \phi_{5/2^+}$(radians) vs $W$(MeV): Phase
  motion obtained with and without $u$-channel terms in Fit~II. 
  Including $u$-channel terms only creates discrepancies in the phase motion
  where one of the $s$-channel waves has a small contribution to the cross 
  section.
  The error bars on both phase motion plots are purely statistical.
}
\label{fig:effects-of-u}
\end{center}
\end{figure}

We can examine what effect adding these terms would have on the resonance
parameters extracted in Fit~II.
Figure~\ref{fig:effects-of-u} shows the phase motion obtained from 
Fit~II with and without $u$-channel amplitudes. The agreement is very good in 
the region where both of the $s$-channel waves have strong contributions to the
cross section. It is only in the regions where the cross section of 
one of the $s$-channel terms is very small that including $u$-channel terms 
leads to a discrepancy in the extracted phase motion. 
Therefore, the conclusions drawn from Fit~II regarding the resonance states
are unaffected by how the $u$-channel terms are modeled.

We also note here that the likelihoods of the fits containing the $u$-channel
terms were worse in all bins. Perhaps this is not surprising since the modified OTL 
terms were obtained assuming the entire backward production amplitude is due to 
$u$-channel mechanisms at our highest energies.
To obtain a better $u$-channel model, the OTL parameters should be fit 
including $s$-channel waves; however, to simply estimate the effects
of neglecting $u$-channel terms in Fit~II, these parameters are sufficient.

\section{\label{section:conc}CONCLUSIONS}

An event-based mass-independent partial wave analysis has been performed on
data obtained using the CLAS at Jefferson Lab. Evidence has been found for contributions from 
the $F_{15}(1680)$ and $D_{13}(1700)$ nucleon resonance states. These states are found to be 
dominant near threshold. The data also strongly support the presence of the $G_{17}(2190)$ state.
Suggestive evidence for an additional $5/2^{+}$ state with a mass around $1.9$--$2$~GeV 
has also been found. The data shows definite strength in this partial wave over a very 
large energy range.  The phase motion between this wave and the $G_{17}(2190)$ supports the
presence of a second $5/2^+$ state near $1.95$~GeV. Some evidence for other states exists, 
although the interpretations are more difficult. The strength seen in the $J^P=3/2^+$ wave around 
$W=1.8-2$~GeV, for example, is not consistent with a single resonant state; however, we cannot 
rule out the possibility that multiple $3/2^{+}$ resonances could be contributing to our data at these 
energies. To extract additional resonance signals from our data, improved theoretical 
input for the non-resonant terms may be required.  In particular, including 
the amplitudes currently being generated by coupled-channel analyses would be
highly desirable.

\begin{acknowledgments}
We thank the staff of the Accelerator and the Physics Divisions at
Thomas Jefferson National Accelerator Facility who made this
experiment possible.  
This work was supported in part by the U.S. Department of Energy
(under grant No. DE-FG02-87ER40315), 
the National Science Foundation,
the Italian Istituto Nazionale di Fisica Nucleare, 
the French Centre National de la Recherche Scientifique, 
the French Commissariat \`{a} l'Energie Atomique, 
the Science and Technology Facilities Council (STFC),
and the Korean Science and Engineering Foundation.  
The Southeastern Universities Research Association (SURA)
operated Jefferson Lab under United States DOE contract
DE-AC05-84ER40150 during this work.
\end{acknowledgments}

\end{document}